\documentclass[aoas,preprint]{imsart}

\RequirePackage[OT1]{fontenc}
\RequirePackage{amsthm,amsmath}
\pdfoutput=1
\RequirePackage{natbib}
\RequirePackage[colorlinks,citecolor=blue,urlcolor=blue]{hyperref}

\usepackage{epsfig, epsf, graphicx, subfigure, epstopdf}
\usepackage{amssymb,amsmath,bm}
\usepackage{natbib}
\usepackage{multirow}
\usepackage{pstricks, pst-node, psfrag}
\usepackage{rotating, lscape}

\def\be{\begin{eqnarray}}
\def\ee{\end{eqnarray}}
\newcommand{\bmu}{\boldsymbol{\mu}}

\arxiv{arXiv:0000.0000}

\startlocaldefs
\numberwithin{equation}{section}
\theoremstyle{plain}

\endlocaldefs

\begin{document}

\begin{frontmatter}
\title{Reducing Storage of Global Wind Ensembles with Stochastic Generators}
\runtitle{Stochastic Wind Generators}

\begin{aug}
\author{\fnms{Jaehong} \snm{Jeong}\thanksref{t1}\ead[label=e1]{jaehong.jeong@kaust.edu.sa}},
\author{\fnms{Stefano} \snm{Castruccio}\thanksref{m2}\ead[label=e2]{scastruc@nd.edu}},
\author{\fnms{Paola} \snm{Crippa}\thanksref{m2}\ead[label=e3]{pcrippa@nd.edu}}
\and
\author{\fnms{Marc} \snm{G. Genton}\thanksref{t1}\ead[label=e4]{marc.genton@kaust.edu.sa}}

\thankstext{t1,m2}{This publication is based upon work supported by the King Abdullah University of Science and Technology (KAUST) Office of Sponsored Research (OSR) under Award No: OSR-2015-CRG4-2640.}
\runauthor{J. Jeong et al.}

\affiliation{King Abdullah University of Science and Technology\thanksmark{t1} and University of Notre Dame\thanksmark{m2}}

\address{Jaehong Jeong and Marc G. Genton\\
CEMSE Division\\
King Abdullah University of Science\\
\ \ and Technology\\
Thuwal, 23955-6900\\ 
Saudi Arabia\\
\printead{e1}\\
\printead{e4}
}

\address{Stefano Castruccio\\
Department of Applied and Computational\\ 
\ \ \ Mathematics and Statistics\\
University of Notre Dame\\
Notre Dame, IN 46556\\
United States of America\\
\printead{e2}
}

\address{Paola Crippa\\
Department of Civil \& Environmental\\
\ \ \ Engineering \& Earth Science\\
University of Notre Dame\\
Notre Dame, IN 46556\\
United States of America\\
\printead{e3}
}

\end{aug}

\begin{abstract}

Wind has the potential to make a significant contribution to future energy resources. Locating the sources of this renewable energy on a global scale is however extremely challenging, given the difficulty to store very large data sets generated by modern computer models. We propose a statistical model that aims at reproducing the data-generating mechanism of an ensemble of runs via a Stochastic Generator (SG) of global annual wind data. We introduce an evolutionary spectrum approach with spatially varying parameters based on large-scale geographical descriptors such as altitude to better account for different regimes across the Earth's orography. We consider a multi-step conditional likelihood approach to estimate the parameters that explicitly accounts for nonstationary features while also balancing memory storage and distributed computation. We apply the proposed model to more than 18 million points of yearly global wind speed. The proposed SG requires orders of magnitude less storage for generating surrogate ensemble members from wind than does creating additional wind fields from the climate model, even if an effective lossy data compression algorithm is applied to the simulation output.

\end{abstract}


\end{frontmatter}

\section{Introduction}\label{sec:intro}

Environmental and societal concerns about climate change are prompting many countries to seek alternative energy resources \citep{moomaw2011renewable,Obamaaam6284}. Wind is a clean and renewable energy source that has the potential to substantially contribute to energy portfolios without causing negative environmental impacts \citep{wiser2011wind} and that can reduce the quantity of anthropogenic greenhouse gases on global warming \citep{barthelmie2014potential}. In order to provide energy assessments in developing countries where no regional studies are available, Earth System Models (ESMs) currently represent a valuable tool to investigate where sustainable wind resources are located. While ESMs are important for physically consistent projections, they represent only an approximation of the true state of the Earth's system, thereby representing uncertainty. In particular, small perturbations in the initial conditions generate a plume of simulations whose uncertainty (internal variability) needs to be quantified. While performing sensitivity analysis from internal variability is a fundamental task, a typical collection (ensemble) of runs, such as the Coupled Model Intercomparison Phase 5 (CMIP5) \citep{taylor2012overview}, comprises a small number of ESM runs, making a detailed assessment infeasible. The Community Earth System Model (CESM) Large ENSemble project (LENS) from the National Center for Atmospheric Research (NCAR) was implemented to provide a large collection of climate model simulations to assess projections in the presence of internal variability with the same forcing scenario \citep{kay2015community}. This ensemble required an enormous effort for only a single scenario (10 million CPU hours and more than 400 terabytes of storage), and very few academic institutions or national research centers have the resources for such an undertaking.

To mitigate storage issues arising when generating such large amounts of data, NCAR has proposed a series of investigations on the topic of reducing storage needs for climate model output. \cite{baker2014methodology} investigated the applicability of lossless and lossy compression algorithms to climate model output. Lossless and lossy compression algorithms respectively provide an exact reconstruction of the data or a reconstruction with some loss of information. \cite{baker2016compression} reported that a lossy algorithm for LENS achieves data reduction that does not impact general scientific conclusions. \cite{guinness2016compression} introduced a compression approach based on a set of summary statistics and a statistical model for the mean and covariance structure in the climate model output. 

Statistical models can provide appropriate stochastic approximations of the spatio-temporal characteristics of the model output, and hence they can be used as surrogates of the original runs \citep{mearns2001wg}. \cite{castruccio2013global}, \cite{castruccio2014beyond}, \cite{castruccio2016compressing}, and \cite{castruccio2017evolutionary} introduced a Stochastic Generator (SG) for annual temperature data to investigate internal variability for different ensembles, assuming that the observed ensemble members were realizations of an underlying statistical model. This approach allowed them to generate runs that were visually indistinguishable from the original model output. In this work, we operate under this framework. 

This work is part of an ongoing collaborative effort with NCAR to develop solutions to deal with memory-intensive models and of a series of investigations sponsored by KAUST to develop novel statistical methodologies to assess wind resources in Saudi Arabia and more broadly in developing countries by relying on ESMs. Various approaches have been proposed to model wind in space and time, see the reviews by \cite{soman2010review} and \cite{zhu2012short}. For LENS, we establish a SG that accounts for the spatio-temporal dependence of the data and uses its parameters to generate additional surrogate runs and efficiently assess the uncertainty in multi-decadal projections. 

Wind fields are expected to exhibit varying spatio-temporal smoothness across longitudes, which is associated with land/ocean regimes and orography. Differences in altitude produce thermal effects as well as acceleration of wind flows over hills, and funneling effects in narrow valleys \citep{banuelos2011methodologies}, and these features are expected to impact the spatial smoothness of this variable. We introduce an evolutionary spectrum approach \citep{priestley1965evolutionary}\footnote{The evolutionary spectrum generalizes the spectrum of a stationary process, by allowing it to vary across longitude while still retaining positive definite covariance functions.}, coupled with spatially varying parameters depending on the surface altitude to better account for different regimes across the Earth's orography. We further introduce a model that allows the latitudinal spectral dependence to vary across different wavenumbers, which markedly improves the fit and allows to model complex latitudinal nonstationarities. 

We perform inference via a multi-step conditional likelihood approach, and we show how the resulting model reduces computational burden and storage costs. Once the parameters are estimated, the proposed model can generate surrogates of ESM runs with different initial conditions within seconds on a modest laptop. The SG requires a small data set of approximately 30 megabytes that describes the mean structure and the parameters of the space-time covariance, whereas downloading a single wind variable from 40 LENS runs requires 1.1 gigabytes.

The remainder of the paper is organized as follows. Section \ref{sec:data} describes the LENS data set. Section \ref{sec:covariance} details the space-time statistical model and the inferential approach. Section \ref{sec:comparison} provides a model comparison and validation of local behavior. Section \ref{sec:simul} illustrates how to generate runs, validate the large scale behavior, and assess the internal variability of global wind fields and wind power densities. The article ends with Section \ref{sec:concl}, which offers a discussion and concluding remarks.

\section{The Large Ensemble}\label{sec:data}

\begin{figure}[htbp]\centering
\includegraphics[height=1.3in,width=2.4in]{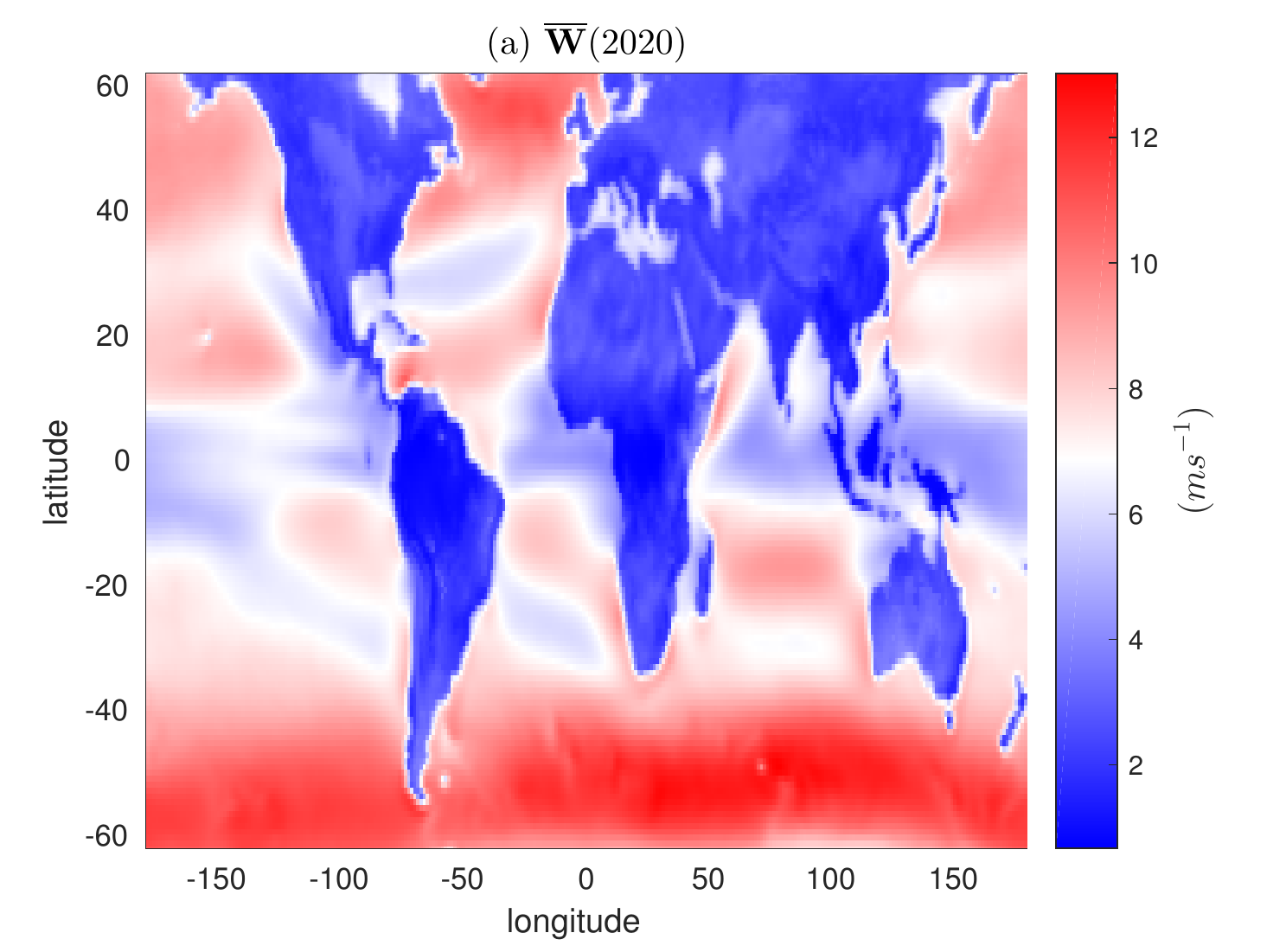}
\includegraphics[height=1.3in,width=2.4in]{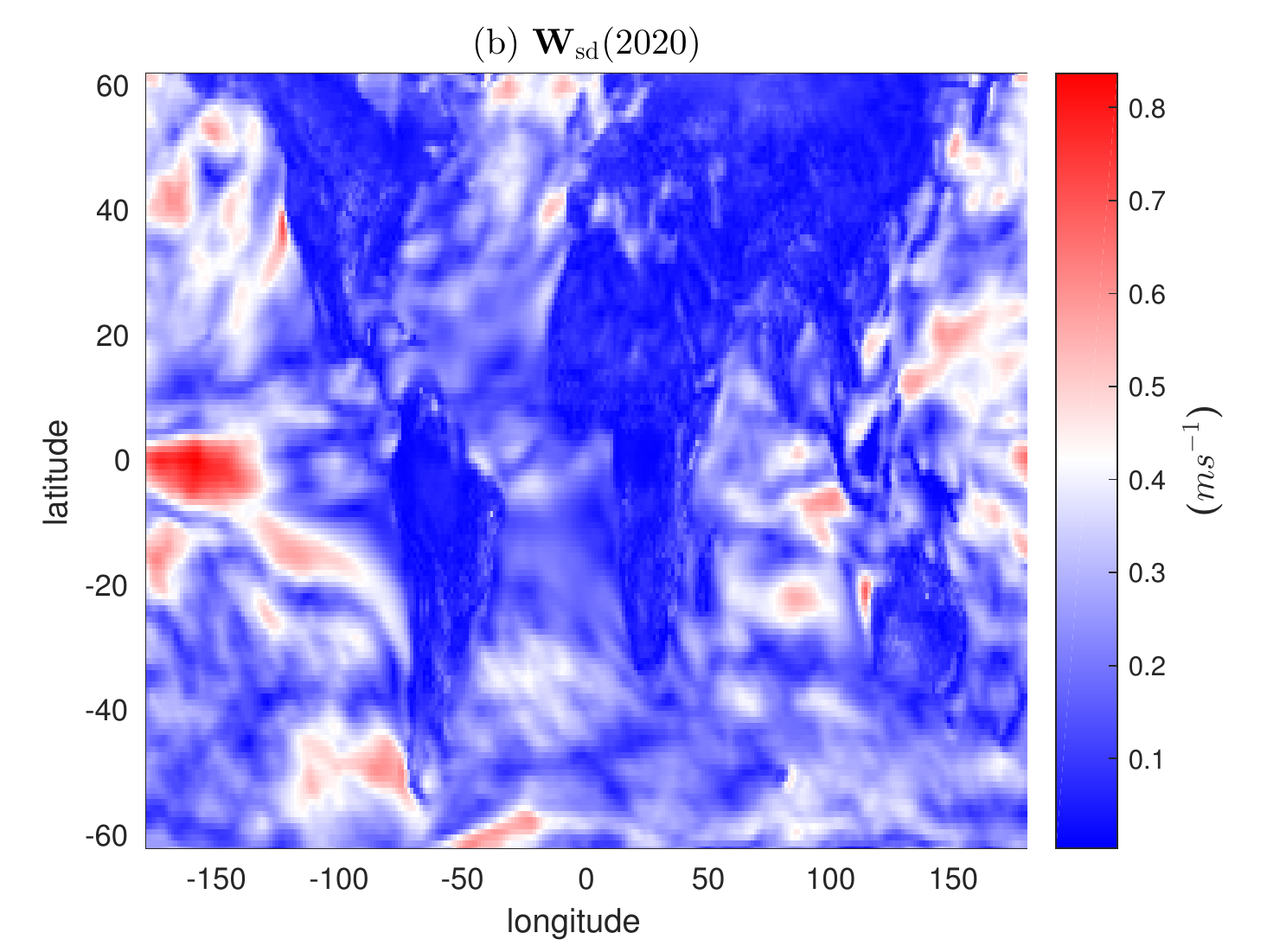}
\caption{The (a) ensemble mean $\overline{\bf W}(2020)=\sum_{r=1}^{R}{\bf W}_{r}(2020)/R$ and (b) ensemble standard deviation ${\bf W}_{\rm sd}(2020)=\sqrt{\sum_{r=1}^{R}\{{\bf W}_{r}(2020)-\overline{\bf W}(2020)\}^{2} /R}$, where $R$ is the number of ensemble members, of the yearly near-surface wind speed (in $m s^{-1}$) for $R=5$.}\label{fig:statistics}
\end{figure}

We focus on LENS, an ensemble of CESM runs with version 5.2 of the Community Atmosphere Model from NCAR \citep{kay2015community}. The ensemble comprises 40 runs of coupled simulations for the period between 1920 and 2100 at $0.9375^{\circ}\times 1.25^{\circ}$ (latitude $\times$ longitude) resolution. Each member is subject to the same radiative forcing scenario: historical up to 2005 and the Representative Concentration Pathway (RCP) 8.5 \citep{vanvuuren11} thereafter. We focus on yearly wind speed at 10~m (computed from the monthly U10 variable) and, since our focus is on future wind trends, we analyze the projections from 2006 to 2100, for a total of 95 years. In the supplementary material (Figure S1, \citep{jeong2017reducingsupplement}), we use a lack of fit index to assess the number of runs $R$ required in the training set for a satisfactory fit, and for this work we establish $R=5$, randomly chosen from the original ensemble. We consider all 288 longitudes, and we discard latitudes near the poles as they would lead to numerical instabilities due to the very close physical distance of neighboring points and the very different statistical behavior of wind speed in the Arctic and Antarctic regions \citep{mcinnes2011global}. We therefore focus on 134 bands between $62^{\circ}$S and $62^{\circ}$N, and the full dataset comprises more than 18 million points ($5 \times 95 \times 134\times 288$). In Figure~\ref{fig:statistics}, we show the ensemble mean and standard deviation of the yearly wind speed from the five chosen runs, in 2020. 

\section{The Space-Time Covariance Model}\label{sec:covariance}

\subsection{A Review of Statistical Models on a Sphere}\label{sec:existingM}

Recently, \cite{gneiting2013strictly} and \cite{ma2015isotropic} provided an overview of isotropic covariance functions for Gaussian processes on a sphere based on geodesic distance. \cite{porcu2016spatio} proposed spatio-temporal covariance and cross-covariance models based on geodesic distance and \cite{clarke2016regularity} studied the regularity properties of Gaussian random fields on a sphere across time. For nonstationary covariance models on a sphere, various construction approaches, such as differential operators \citep{jun2007approach,jun2008nonstationary,jun2011non,jun2014nonstationary}, spherical harmonic representation \citep{stein2007spatial,hitczenko2012some}, stochastic partial differential equations \citep{lindgren2011explicit,bolin2011spatial}, kernel convolution \citep{heaton2014constructing} and deformation \citep{das00} have been introduced. A new review of spherical process models for global spatial statistics can be found in \citep{jjg17}.

When modeling global data, a common assumption is that the (Gaussian) spatial process is {\it axially symmetric}, i.e., its mean depends on latitude, $L$, and its covariance depends only on the longitudinal lag, $\ell_{1}-\ell_{2}$, between two points \citep{jones1963stochastic}. This class of models implies that data are stationary at a given latitude, but this assumption is clearly inappropriate for many variables whose dynamics are influenced by the presence of large-scale geographical descriptors such as land and ocean. To better account for different statistical characteristics of variables such as temperature or wind speed, more flexible nonstationary models are needed. \cite{jun2014nonstationary} considered nonstationary models with a differential operator approach and spatially varying smoothness parameters over land and ocean. \cite{castruccio2017evolutionary} also relaxed the assumption of axial symmetry by proposing an evolutionary spectrum approach to account for different regimes over land and ocean. In this work, we propose a generalization of this approach to allow spatial smoothness to change with orography, and a novel approach for changing spectral dependence across latitudes for different wavenumbers.

\subsection{The Statistical Framework}\label{sec:newM}

Climate model variables in the atmospheric component tend to forget their initial conditions after a small number of time steps. Each ensemble member evolves in `deterministically chaotic' patterns after the climate model forgets its initial state \citep{lorenz1963deterministic}. \cite{collins2002climate}, \cite{collins2002assessing}, and \cite{branstator2010two} discussed the validity of the deterministically chaotic nature of climate models. Since ensemble members from the LENS differ only in their initial conditions \citep{kay2015community}, each one will be treated as a statistical realization from a common Gaussian distribution (see Figure S2 for two normality tests for this data set). Denote by ${W}_{r}(L_{m},\ell_{n},t_{k})$ the spatio-temporal near-surface wind speed for realization $r$ at the latitude $L_{m}$, longitude $\ell_{n}$, and time $t_{k}$, where $r=1,\dots,R$, $m=1,\dots,M$, $n=1,\dots,N$, and $k=1,\dots,K$. Define the vector
\be
{\bf W}_{r}=\{{W}_{r}(L_{1},\ell_{1},t_{1}),\dots,{W}_{r}(L_{M},\ell_{1},t_{1}),{W}_{r}(L_{1},\ell_{2},t_{1}),\dots,{W}_{r}(L_{M},\ell_{N},t_{K})  \}^{\top}. \nonumber
\ee
We assume that ${\bf W}_{r}$ is independent across $r$ conditional on its climate: 
\be\label{mod_ensemble}
{\bf W}_{r}=\bmu+{\bm \epsilon}_{r},\ \ {\bm \epsilon}_{r}\stackrel{\rm iid}{\sim}\mathcal{N}({\bm 0},{\bm \Sigma}({\bm \theta})),\label{eq:model}
\ee
where ${\bm \mu}$ is the space-time mean across realizations and ${\bm \theta}$ is a vector of fixed and unknown covariance parameters. By assuming independent realizations, we can estimate ${\bm \theta}$ using a restricted log-likelihood without providing any parametrization of ${\bm \mu}$. \cite{castruccio2013global} provided the following expression for twice the negative restricted log-likelihood function: 
\be
\begin{array}{lll}
2l({\bm\theta};{\bf D}) & = & KNM(R-1)\log(2\pi)+KNM\log(R)\\
                                  && +(R-1)\log|{\bm \Sigma}({\bm\theta})|+\sum_{r=1}^{R}{\bf D}_{r}^{\top} {\bm \Sigma}({\bm\theta})^{-1}{\bf D}_{r},\label{eq:likelihood1}
\end{array}
\ee
where ${\bf D}=({\bf D}_{1}^{\top},\dots,{\bf D}_{R}^{\top})^{\top}$ and ${\bf D}_{r}={\bf W}_{r}-\overline{\bf W}$ where $\overline{\bf W}=\sum_{r=1}^{R}{\bf W}_{r}/R$. We use this expression throughout this work.

\subsection{Temporal Dependence}\label{first_step}

Let ${\bm \epsilon}_{r}(t_{k})$ be the vector of the stochastic component of \eqref{eq:model} for time $t_{k}$ and realization $r$. No evidence of nonstationarity in time was found, and we assume a Vector AutoRegressive of order 2 (VAR(2)) structure for ${\bm \epsilon}_{r}(t_{k}),$ with different parameters for each spatial location. Diagnostics showed no evidence of the need for higher order autoregressive coefficients or cross-temporal dependence (Figures S3 and S4 in the supplementary material \citep{jeong2017reducingsupplement}). A non-negligible temporal dependence across locations (as observed at higher temporal resolutions) would imply a nonseparable model. Our model can be modified to allow for interactions of temporal dependence across neighboring locations \citep{tagle2017}. The VAR(2) model is
\be\label{eq:ar2}
{\bm \epsilon}_{r}(t_{k})={\bm \Phi}_{1}{\bm \epsilon}_{r}(t_{k-1})+{\bm \Phi}_{2}{\bm \epsilon}_{r}(t_{k-2})+{\bf SH}_{r}(t_{k}), 
\ee
where ${\bm \Phi}_{1}=\text{diag}\{\phi_{L_{m},\ell_{n}}^{1}\}$ and ${\bm \Phi}_{2}=\text{diag}\{\phi_{L_{m},\ell_{n}}^{2}\}$ are two $MN\times MN$ diagonal matrices with autoregressive coefficients, and ${\bf S}=\text{diag}\{S_{L_{m},\ell_{n}}^{1}\}$ is an $MN\times MN$ diagonal matrix with the associated standard deviations, so that the temporal parameters are denoted by $\bm{\theta}_{\rm time}=(\phi^{1}_{L_{m},\ell_n},\phi^{2}_{L_{m},\ell_n},S_{L_{m},\ell_n})^{\top}$ for $n=1,\ldots,N$ and $m=1,\ldots,M$. For all spatial locations, we estimate ${\bm \Phi}_{1}$, ${\bm  \Phi}_{2}$, and ${\bf S}$ by assuming that the innovations ${\bf H}_{r}(t_{k})=\{{H}_{r}(L_{m},\ell_{n},t_{k})\}$ are independent across latitude and longitude. This allows us to perform inference in parallel: each spatial location can be estimated independently by a core in a workstation or cluster. Here, ${\bf H}_{r}(t_{k})\stackrel{\rm iid}{\sim}\mathcal{N}({\bm 0},{\bm C})$, and the following Sections \ref{second_step} and \ref{third_step} are entirely devoted to determining the ${\bf H}_{r}(t_{k})$ for ${\bm C}$.

\begin{figure}[!ht]\centering
\includegraphics[height=1in,width=1.6in]{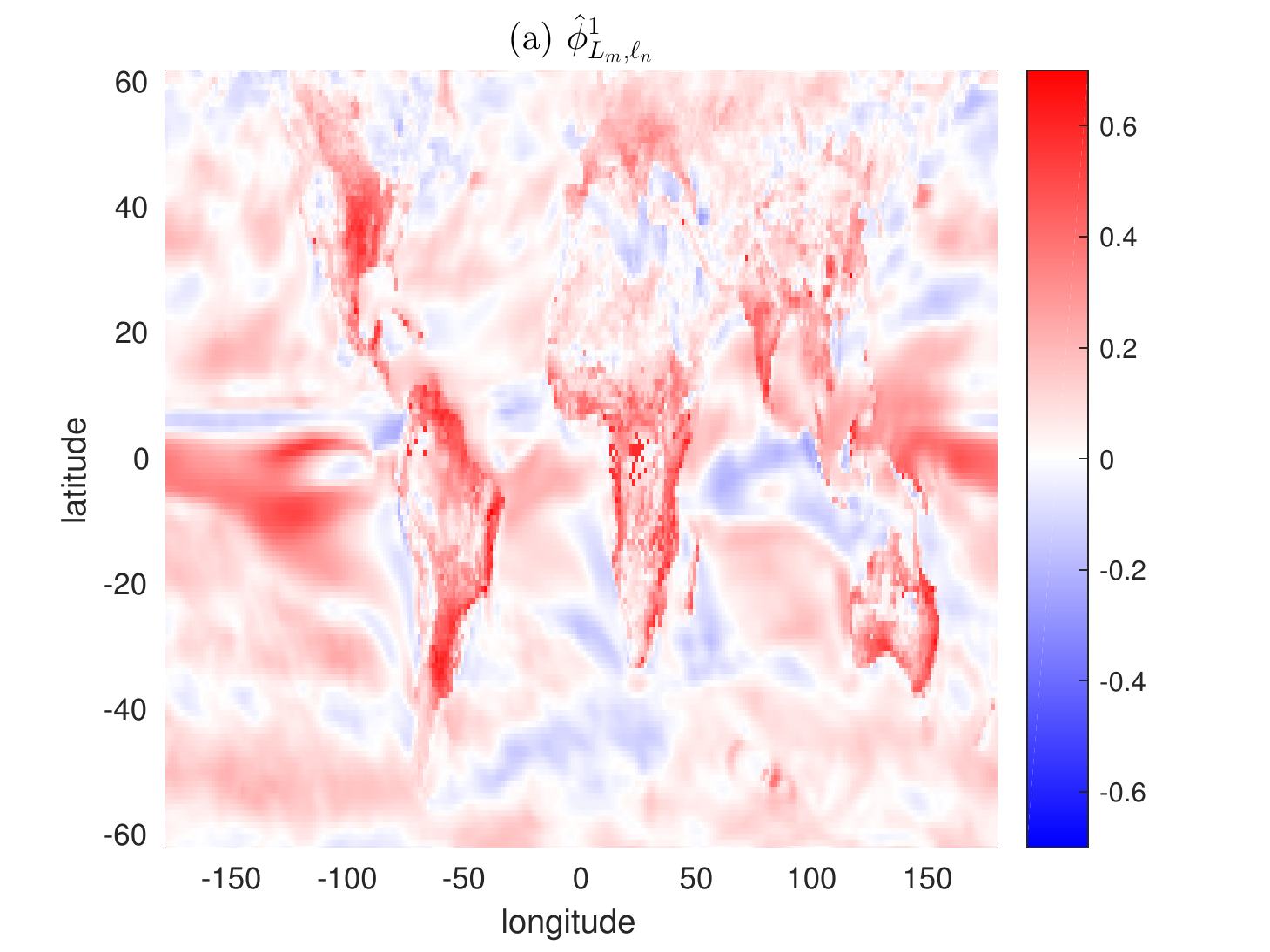}
\includegraphics[height=1in,width=1.6in]{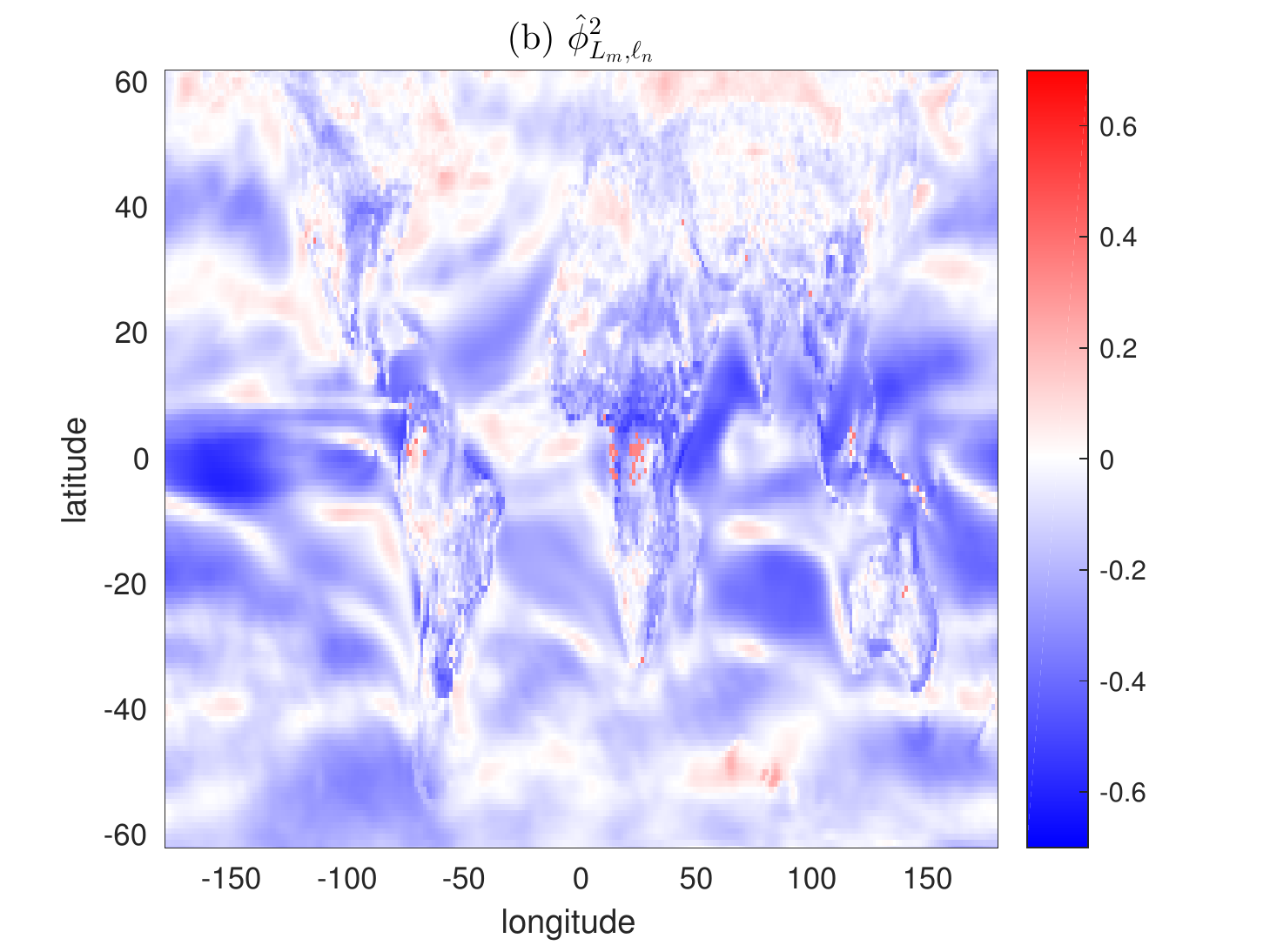}
\includegraphics[height=1in,width=1.6in]{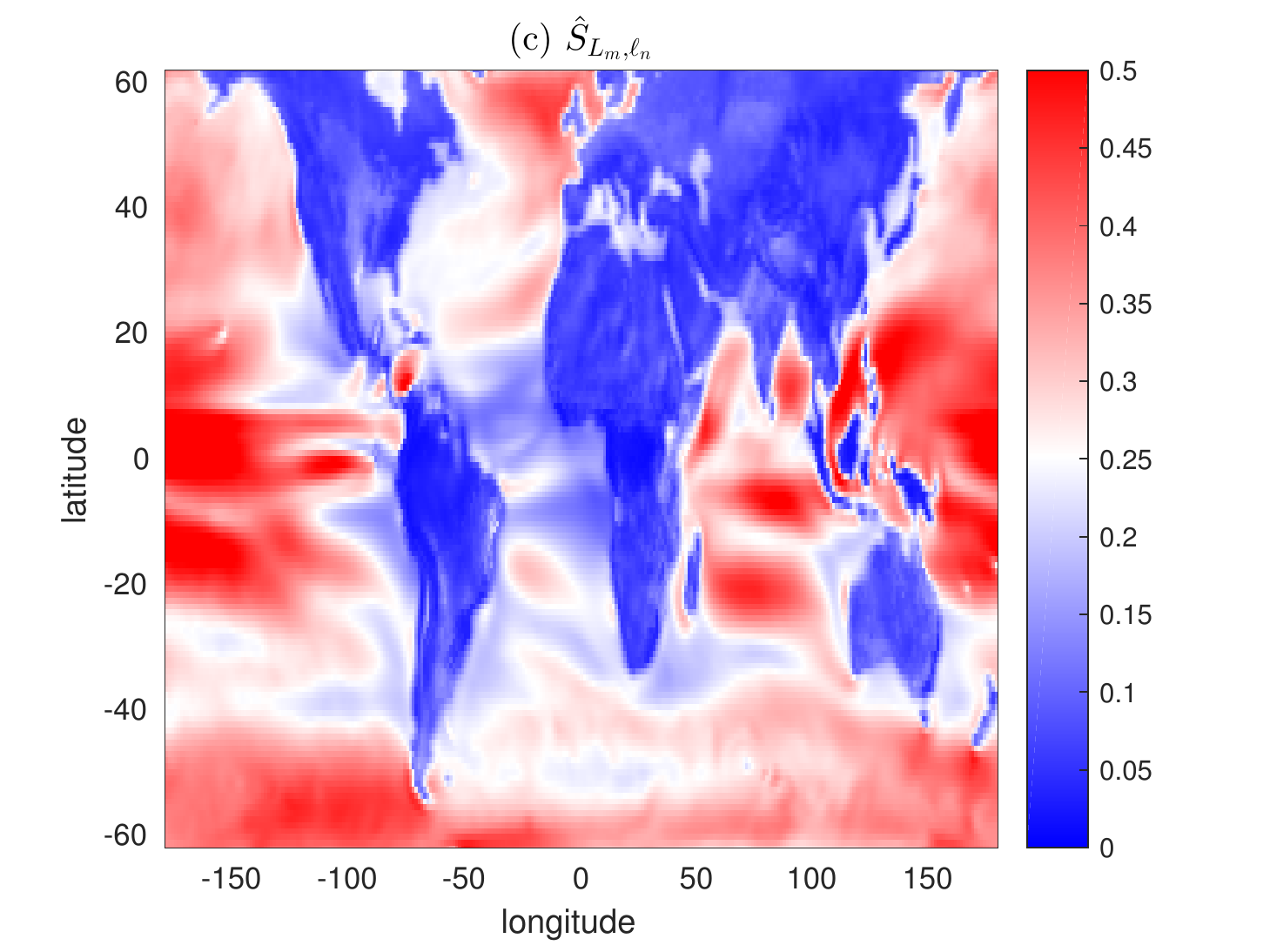}
\caption{Plots of the estimated autoregressive parameters for the temporal model as defined in \eqref{eq:ar2}: (a) $\hat{\phi}_{L_{m},\ell_{n}}^{1}$, (b) $\hat{\phi}_{L_{m},\ell_{n}}^{2}$, and (c) $\hat{S}_{L_{m},\ell_{n}}$.}\label{fig:AR2}
\end{figure}

Figure~\ref{fig:AR2} shows the estimated autoregressive parameters. The two autoregressive coefficients, $\phi_{L_{m},\ell_{n}}^{1}$ and $\phi_{L_{m},\ell_{n}}^{2}$, are estimated to be mostly positive and negative, respectively (corresponding $p$-values are available in Figure~S3 in the supplementary material \citep{jeong2017reducingsupplement}). $\hat{S}_{L_{m},\ell_{n}}$ exhibits higher values over ocean than over land. The marginal standard deviation shows similar patterns to $\hat{S}_{L_{m},\ell_{n}}$ with a different scale (not shown).

\subsection{Longitudinal Dependence}\label{second_step}

We now provide a model for the spatial correlation of the unscaled innovations, ${H}_{r}(L_{m},\ell_{n},t_{k})$, at different longitudes but at the same latitude. An evolutionary spectrum allows for changing behavior across large-scale geographical descriptors. \cite{castruccio2017evolutionary} proposed to model ${H}_{r}(L_{m},\ell_{n},t_{k})$ in the spectral domain by performing a generalized Fourier transform across longitude: 
\be
{H}_{r}(L_{m},\ell_{n},t_{k})=\sum_{c=0}^{N-1}f_{L_{m},\ell_{n}}(c)\exp({i\ell_{n}c})\widetilde{H}_{r}(c,L_{m},t_{k}), \label{eq:gft}
\ee
where $i$ is the imaginary unit, $c=0, \ldots, N-1$ is the wavenumber, $f_{L_{m},\ell_{n}}(c)$ is the evolutionary spectrum across longitude, and $\widetilde{H}_{r}(c,L_{m},t_{k})$ is the transformed process in the spectral domain. 

In this work, we propose a flexible model in which ocean, land, and high mountains with altitude information are included as covariates to better account for the statistical behavior of wind speed. The United Nations Environmental Programme does not provide an unambiguous definition of `mountainous environment' \citep{blyth2002mountain}. Hence, we subjectively choose a threshold value of 1,000~m (see Figure S5 in the supplementary material \citep{jeong2017reducingsupplement} for the global distribution of high mountains). We allow $f_{L_{m},\ell_{n}}(c)$ to depend on $\ell_{n}$ in a land, ocean and high mountain domain so that it can be expressed as
\be
&&f_{L_{m},\ell_{n}}(c)=f_{L_{m},\ell_{n}}^{1}(c)I_{\rm land\cap hmt}(L_{m},\ell_{n})+f_{L_{m},\ell_{n}}^{2}(c)b_{\rm land\cap hmt^{c}}(L_{m},\ell_{n};g_{L_{m}},r_{L_{m}})\nonumber\\
&&\hspace{2cm}+f_{L_{m},\ell_{n}}^{3}(c)\{1-b_{\rm land}(L_{m},\ell_{n};g_{L_{m}},r_{L_{m}})\}, \label{eq:spectra} \\
&&b_{\rm land}(L_{m},\ell_{n};g_{L_{m}},r_{L_{m}})=\sum_{n'=1}^{N}\tilde{I}_{\rm land}(L_{m},\ell_{n};g_{L_{m}})w(L_{m},\ell_{n}-\ell_{n'};r_{L_{m}}),\nonumber
\ee
where $I_{\rm land\cap hmt}(L_{m},\ell_{n})$ is the indicator function for high mountains. The transition between non-mountainous land and ocean in the second and third terms requires a parametrization for a smooth transition. Here, the modified indicator function of $I_{\rm land}(L_{m},\ell_{n})$ is $\tilde{I}_{\rm land}(L_{m},\ell_{n};g_{L_{m}})$, which is equal to 1 for $g_{L_{m}}$ grid points wherever there is a land/ocean transition (this parameter can also be negative) and $w(L_{m},\ell_{n}-\ell_{n'};r_{L_{m}})$ is the Tukey taper function \citep{tukey1967introduction} with range $r_{L_{m}}$ (other taper functions are equally effective). Hence, $b_{\rm land}(L_{m},\ell_{n};g_{L_{m}},r_{L_{m}})$ allows for a smoother transition between land/ocean states by convolving the modified land/ocean indicator, $\tilde{I}_{\rm land}(L_{m},\ell_{n};g_{L_{m}})$, with the taper function, $w(L_{m},\ell_{n}-\ell_{n'};r_{L_{m}})$. We additionally use the information of the surface altitude, which has an impact on land and high mountains. The component spectra in \eqref{eq:spectra} is defined according to the parametric form \citep{castruccio2013global,poppick2014using}:
\be
|f_{L_{m},\ell_{n}}^{j}(c)|^{2}=\phi_{L_{m},\ell_{n}}^{j}\{(\alpha^{j}_{L_{m},\ell_{n}})^{2}+4\sin^{2}(c\pi/N) \}^{\nu_{L_{m},\ell_{n}}^{j}+1/2}, & j=1,2,3,\nonumber
\ee
where $(\phi_{L_{m},\ell_{n}}^{j},\alpha_{L_{m},\ell_{n}}^{j},\nu_{L_{m},\ell_{n}}^{j})$ have a similar interpretation as the variance, inverse range, and smoothness parameters, respectively, for the Mat\'{e}rn spectrum over the line. We allow spatially varying parameters to depend on the surface altitude, with log-linear parametrization to ensure positivity for $\phi_{L_{m},\ell_{n}}^{j}=\beta_{L_{m}}^{j,\phi}\exp[ \tan^{-1}\{A_{L_{m},\ell_{n}} \gamma^{\phi}_{L_{m}}\}]$, $j=1,2$ and $\phi_{L_{m},\ell_{n}}^{3}=\beta_{L_{m}}^{3,\phi}$, where $\beta_{L_{m}}^{j,\phi}$ is a positive number, $\gamma^{\phi}_{L_{m}}$ is a real number, and $A_{L_{m},\ell_{n}}$ represents the altitude at location $(L_{m},\ell_{n})$. $\nu_{L_{m},\ell_{n}}^{j}$ and $\alpha_{L_{m},\ell_{n}}^{j}$ have a similar structure. In order to avoid overparametrization, $\gamma^{\phi}_{L_{m}}$ controls the impact of the surface altitude for land and high mountains, i.e., $\phi_{L_{m},\ell_{n}}^{1}(c)$ and $\phi_{L_{m},\ell_{n}}^{2}(c)$ share the same coefficient, $\gamma^{\phi}_{L_{m}}$. Hence, the longitudinal parameters are ${\bm \theta}_{\rm lon}=(\beta_{L_{m}}^{j,\phi},\gamma^{\phi}_{L_{m}},\beta_{L_{m}}^{j,\nu},\gamma^{\nu}_{L_{m}},\beta_{L_{m}}^{j,\alpha},\gamma^{\alpha}_{L_{m}},g_{L_{m}},r_{L_{m}})^{\top}$, $j=1,2,3$ and $m=1,\ldots,M$. The parameter values for each $L_m$ are independent from the other latitudinal bands, therefore each core of a workstation or cluster can perform inference independently on each band.

\begin{figure}[b!]\centering
\includegraphics[height=1.3in,width=2.4in]{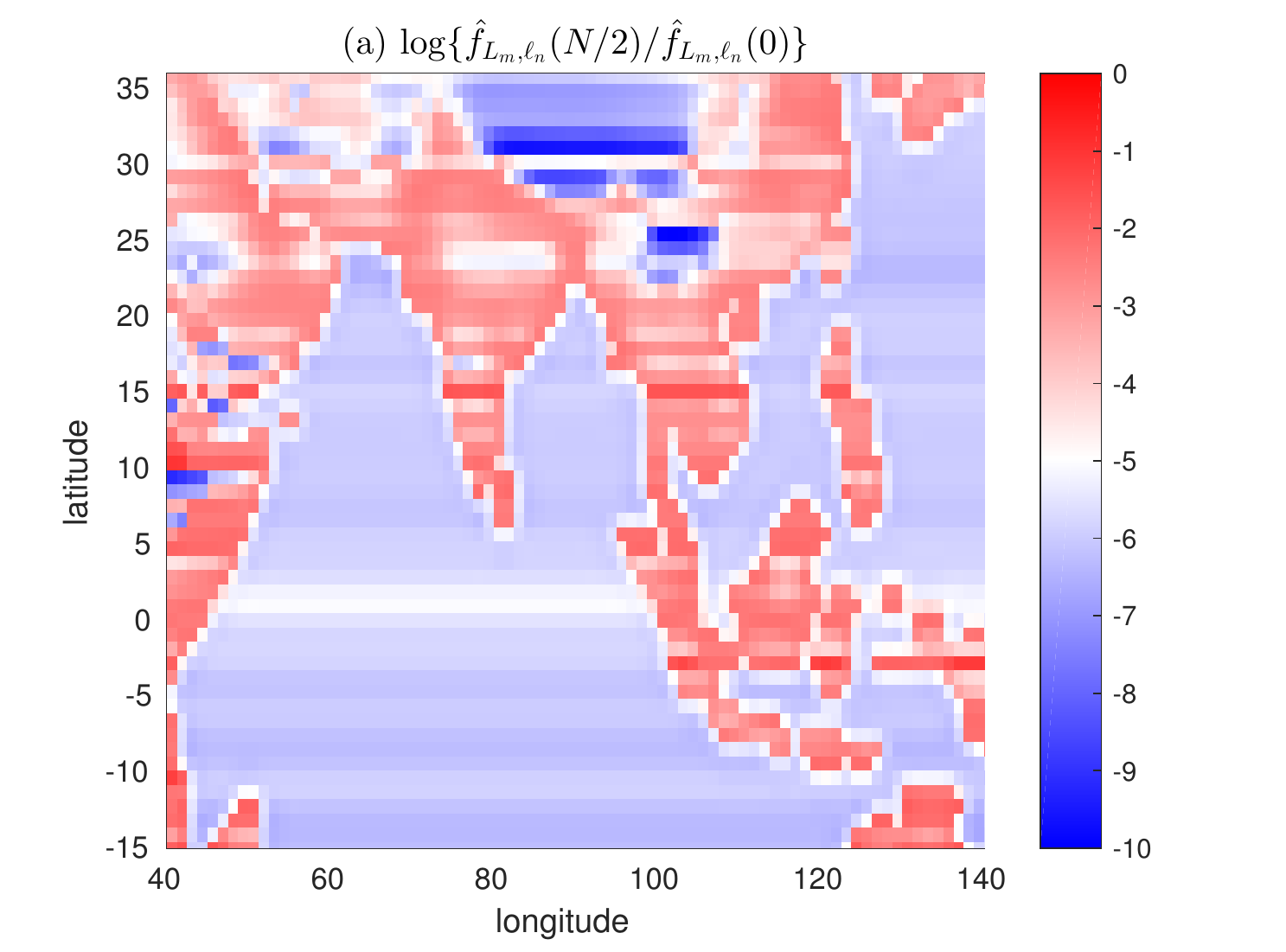}\hspace{-0.4cm}
\includegraphics[height=1.3in,width=2.4in]{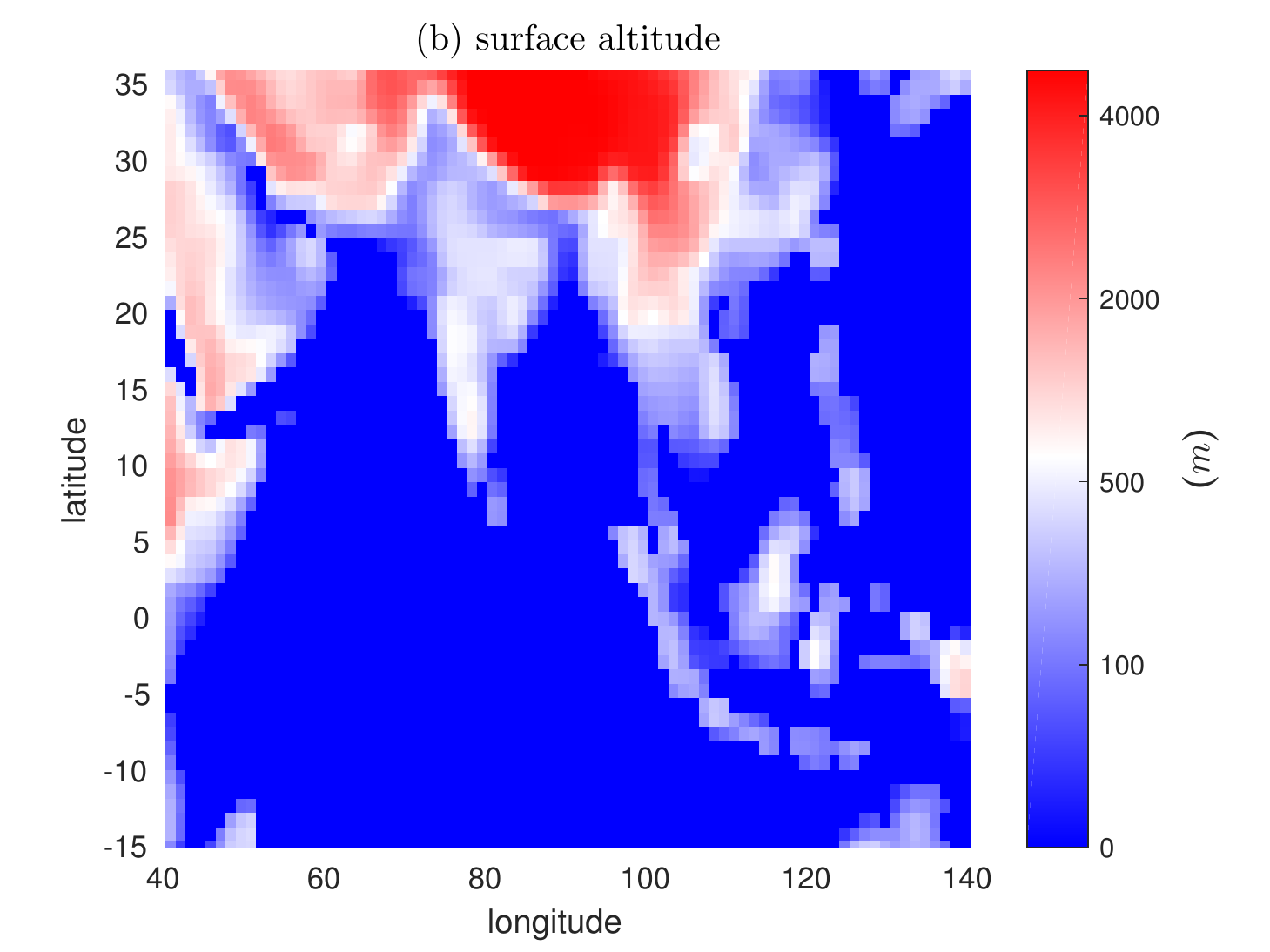}
\caption{(a) Log-ratio of periodograms, $\log\{\hat{f}_{L_{m},\ell_{n}}(N/2)/\hat{f}_{L_{m},\ell_{n}}(0)\}$, and (b) surface altitude (orography) near the Indian Ocean and Himalayan region.}\label{fig:est_spectrum}
\end{figure}

\begin{sloppypar}
In Figure~\ref{fig:est_spectrum}(a), we show $\log\{\hat{f}_{L_{m},\ell_{n}}(N/2)/\hat{f}_{L_{m},\ell_{n}}(0)\}$, the log-ratio of periodograms that empirically estimates the rate of spectral decay at high frequency, and the surface altitude near the Indian Ocean and Himalayan region. At high altitudes, the Himalayan region and Western China exhibit pronounced spectral decay compared to neighboring land masses at low altitudes, such as India and Eastern China. Moreover, the patterns of spectral decay markedly follow the topographical relief, as apparent from Figure~\ref{fig:est_spectrum}(b). Indeed, besides a smoother ocean behavior, annual winds are considerably smoother at high altitudes, as demonstrated by the fast rate of spectral decay over the region corresponding to the Himalayas. 
\end{sloppypar}

\begin{figure}[htbp]\centering
\includegraphics[height=1.4in,width=2.6in]{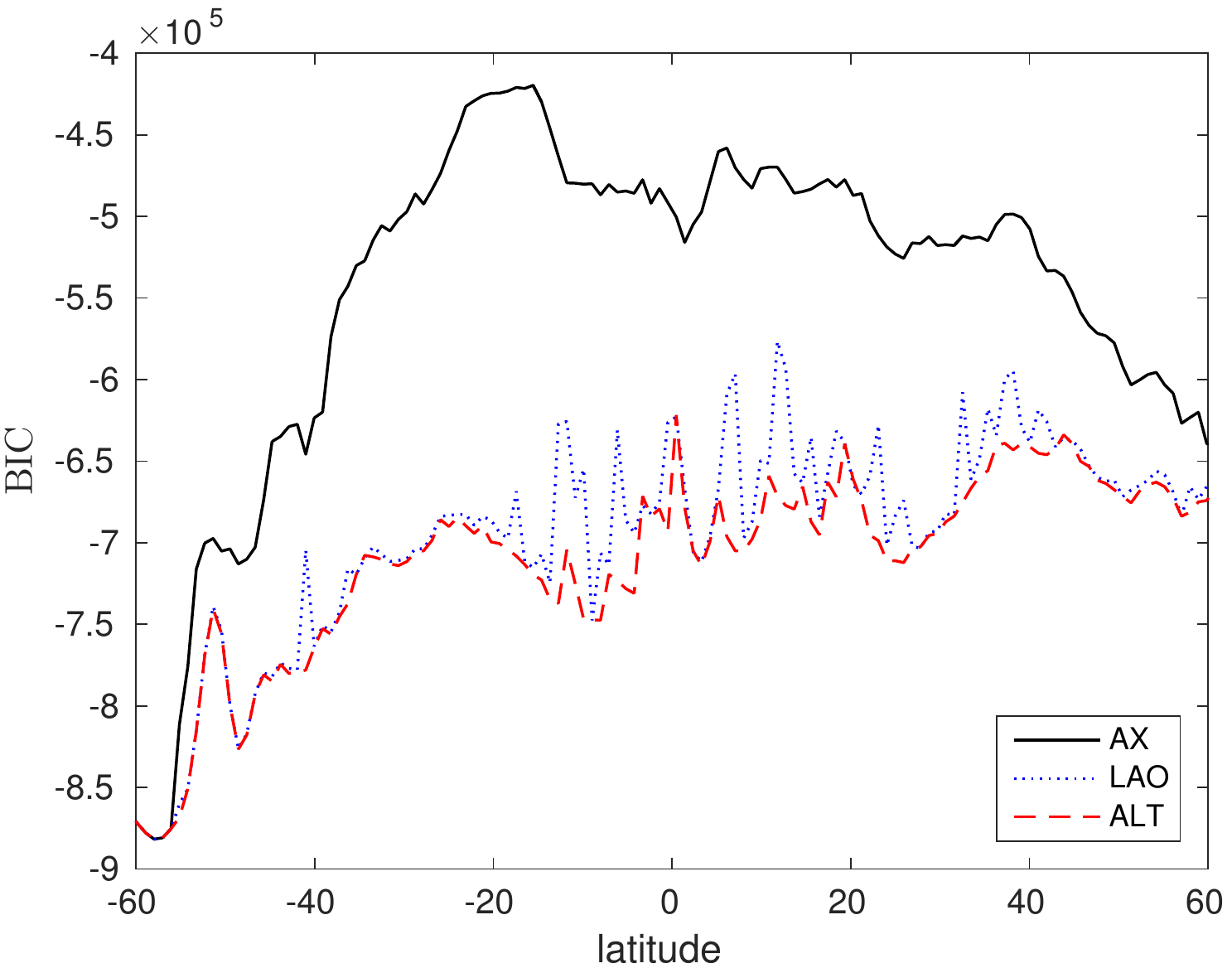}
\caption{Comparison of AX, LAO, and ALT models in terms of BIC versus latitude.}\label{fig:BIC_single}
\end{figure}

Figure~\ref{fig:BIC_single} presents a comparison of three models: the axially symmetric model (AX), the evolutionary spectrum model with land and ocean (LAO), and the new evolutionary spectrum model with altitude (ALT), in terms of the Bayesian Information Criterion (BIC) against latitude. LAO and ALT uniformly outperform AX, but ALT is significantly more flexible than LAO at latitudinal bands between $25^{\circ}$S and $45^{\circ}$N, where the percentage of points with high mountains within these bands is $7.6\%$, compared to $3\%$ within the other bands. 

\subsection{Latitudinal Dependence}\label{third_step}

We propose a novel Vector AutoRegressive model of order 1, VAR(1), across latitudes to allow for dependence of $\widetilde{H}_{r}(c,L_{m},t_{k})$ across neighboring wavenumbers. For any $r$ and $t_{k}$, denote by $\widetilde{\bold{H}}_{L_{m}}=\{ \widetilde{{H}}_{L_{m}}(c_{1}),\dots,\widetilde{{H}}_{L_{m}}(c_{N}) \}^{\top}$, then
\be
\widetilde{\bold{H}}_{L_{m}}&=&\Bigg\{ \begin{array}{c}
\bm{\varphi}_{L_{m}}\widetilde{\bold{H}}_{L_{m-1}}+{\bold{e}}_{L_{m}},\ \ m=2,\dots,M,\\
{\bold{e}}_{L_{1}}\sim\mathcal{N}(\bold{0},\bold{I}),\ \ m=1, \end{array}\label{eq:var1}\\
{\bold{e}}_{L_{m}}&\stackrel{\rm iid}{\sim}&\mathcal{N}(\bold{0},{\bold{\Sigma}_{L_{m}}}),\ \ m>1,\nonumber
\ee
where $\bm{\varphi}_{L_{m}}$ is an $N \times N$ matrix describing the autoregressive coefficients and $\bold{\Sigma}_{L_{m}}$ in an $N \times N$ matrix with the covariance structure of the innovation. We propose the following banded structure, which eases the computational burden by inducing sparsity and also results in a diagonally dominant matrix:
\be\label{phisigma_var}
\hspace{1cm}
\bm{\varphi}_{L_{m}}\resizebox{.8\hsize}{!}{$ {=\left( \begin{array}{cccccccc}
\varphi_{L_{m}}(c_1) & \frac{ \{1-\varphi_{L_{m}}(c_1)\}a_{L_{m}} }{ 4 } & \frac{ \{1-\varphi_{L_{m}}(c_1)\}b_{L_{m}} }{ 4 } & 0 &  \cdots & 0 & 0 & 0\\
\frac{  \{1-\varphi_{L_{m}}(c_2)\}a_{L_{m}} }{ 4 } & \varphi_{L_{m}}(c_2) &  \frac{ \{1-\varphi_{L_{m}}(c_2)\}a_{L_{m}} }{ 4 } & \frac{ \{1-\varphi_{L_{m}}(c_2)\}b_{L_{m}} }{ 4 } &\cdots & 0 & 0 & 0\\
\frac{ \{1-\varphi_{L_{m}}(c_3)\}b_{L_{m}} }{ 4 } &  \frac{ \{1-\varphi_{L_{m}}(c_3)\}a_{L_{m}} }{ 4 } & \varphi_{L_{m}}(c_3) & \frac{ \{1-\varphi_{L_{m}}(c_3)\}a_{L_{m}} }{ 4 } &  \cdots & 0 & 0 & 0\\
\vdots& \vdots & \vdots & \vdots  & \ddots & \vdots & \vdots & \vdots \\
0 & 0 & 0 & 0 & \cdots & \frac{ \{1-\varphi_{L_{m}}(c_{N-1})\}a_{L_{m}} }{ 4 } & \varphi_{L_{m}}(c_{N-1}) & \frac{ \{1-\varphi_{L_{m}}(c_{N-1})\}a_{L_{m}} }{ 4 }\\
0 & 0 & 0 & 0 & \cdots & \frac{ \{1-\varphi_{L_{m}}(c_{N})\}b_{L_{m}} }{ 4 } & \frac{ \{1-\varphi_{L_{m}}(c_{N})\}a_{L_{m}} }{ 4 } & \varphi_{L_{m}}(c_N)\end{array} \right)}, $}
\ee
where  $a_{L_{m}}, b_{L_{m}} \in (-1,1)$ for all $m$, $\bold{\Sigma}_{L_{m}}=\text{diag}\{1-\varphi_{L_{m}}(c_n)^2\}$ and
\be\label{eq_coh}
\varphi_{L_{m}}(c)=\frac{\xi_{L_{m}}}{\{1+4\sin^{2}(c\pi/N) \}^{\tau_{L_{m}}}},
\ee
where $\xi_{L_{m}}\in [0,1]$ and $\tau_{L_{m}}>0$ for all $m$. If $a_{L_{m}}=b_{L_{m}}=0$, this model corresponds to a nonstationary AR(1) process in latitude:
\be
{\rm corr}\{\widetilde{H}_{r}(c,L_{m},t_{k}),\widetilde{H}_{r'}(c',L_{m'},t_{k'}) \}={\bf 1}\{c=c',k=k',r=r' \}\rho_{L_{m},L_{m'}}(c), \nonumber
\ee
where $\rho_{L_{m},L_{m'}}(c)=\prod_{j=m}^{m'} \varphi_{L_{j}}(c), m<m'$ is the coherence between latitudes $L_{m}$ and $L_{m'}$ among the $\widetilde{H}_{r}(c,L_{m},t_{k})$s with the same wavenumber, time, and realization \citep{castruccio2017evolutionary}. 

\begin{figure}[htbp]\centering
\includegraphics[height=1.3in,width=2.4in]{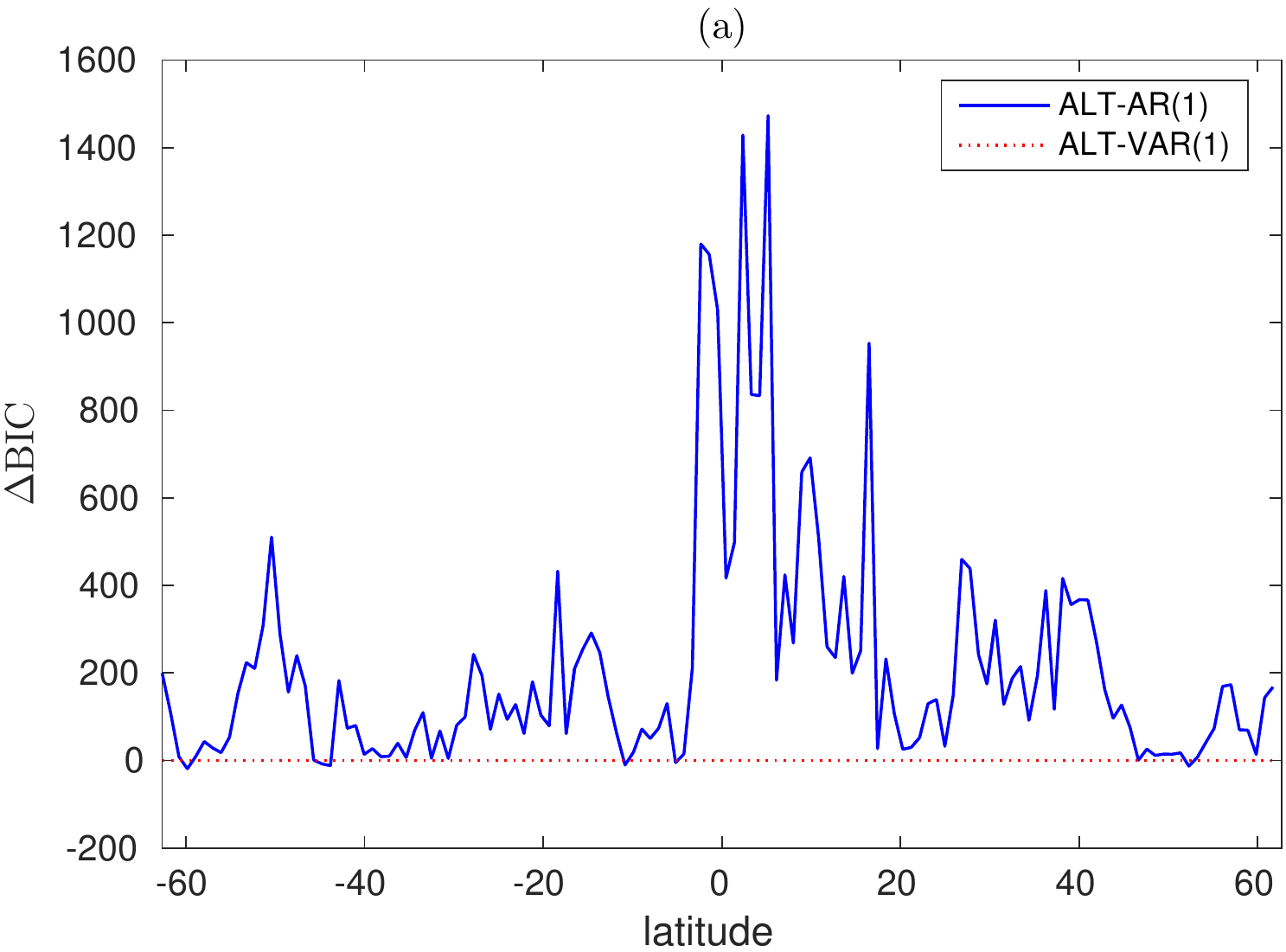}
\includegraphics[height=1.3in,width=2.4in]{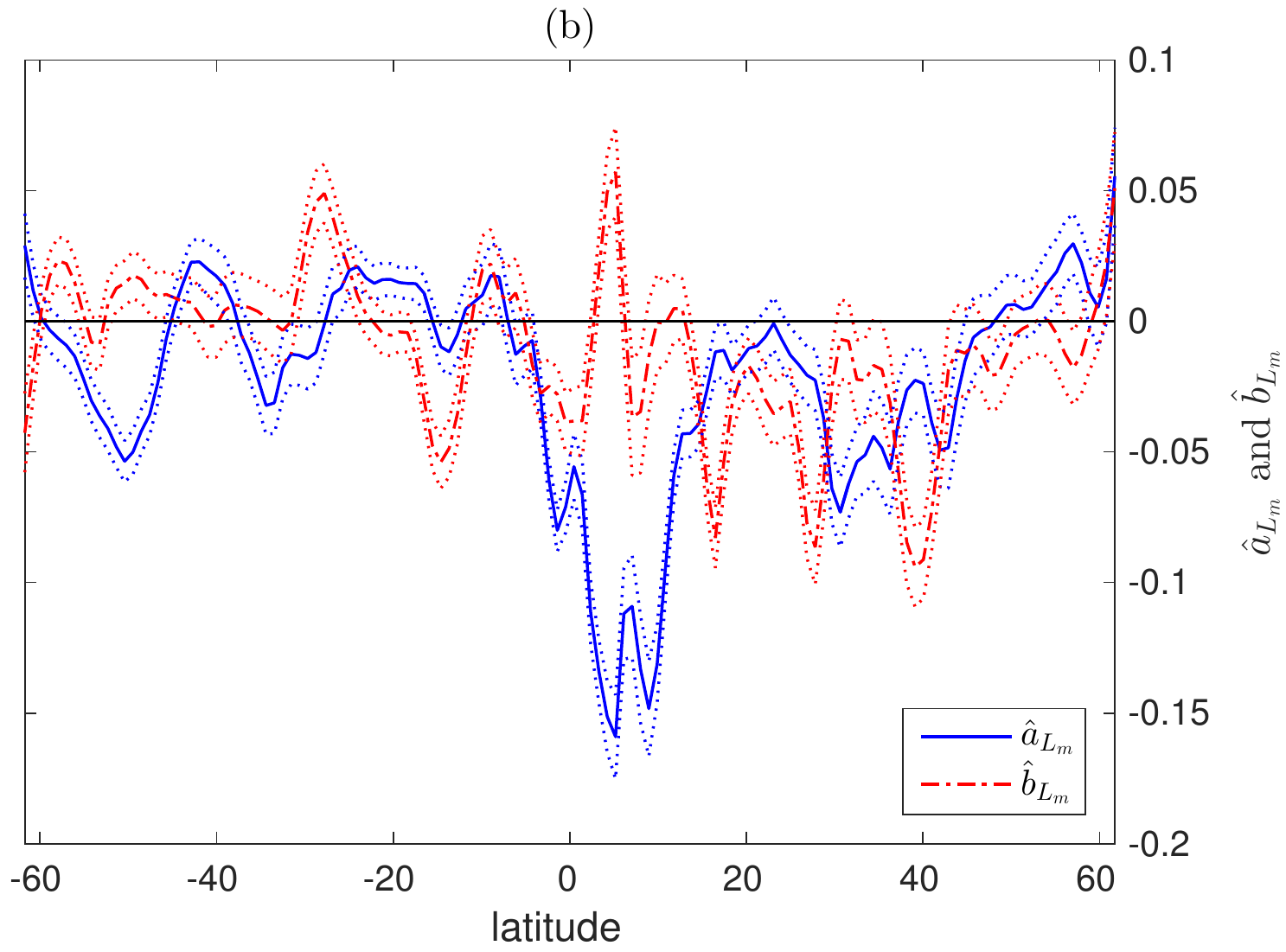}
\caption{Comparison between AR(1) and VAR(1) latitudinal models for adjacent bands in terms of (a) BIC and (b) $\hat{a}_{L_{m}}$ and $\hat{b}_{L_{m}}$ as in \eqref{phisigma_var} (the dotted lines represent the $95\%$ confidence bands). A smoothing spline has been applied to the parameters estimated in (b).}\label{fig:var1}
\end{figure}

To compare VAR(1) with AR(1), we perform inference for every pair of contiguous bands ($L_{m},L_{m+1}$) independently for both models, and we report the BIC and parameter estimates in Figures~\ref{fig:var1}(a) and (b), respectively. For most latitudes, VAR(1) has a large BIC improvement compared with AR(1), and $\hat{a}_{L_{m}}$ and $\hat{b}_{L_{m}}$ are significantly different from 0 (see confidence bands).

To complete the model, the latitudinal dependence of $a_{L_{m}}, b_{L_{m}}$ in \eqref{phisigma_var} and $\xi_{L_{m}}, \tau_{L_{m}}$ in \eqref{eq_coh} must be specified. Figure \ref{fig:var1}(b) highlights how latitudes near the equator result in $\hat{a}_{L_m}$ and $\hat{b}_{L_m}$ being considerably (and significantly) different from zero, hence the need of different coefficients near these latitudinal bands. To mitigate, however, the increased computational cost derived from these additional parameters we choose the bounds $-30^{\circ}$ and $30^{\circ}$, consistently with \cite{castruccio2017evolutionary}, in order to include the tropics, whose climate is determined by the complex interactions between large-scale atmospheric circulation, atmospheric convection, solar and terrestrial radiactive transfer, boundary layers, and clouds \citep{betts1988coupling}. As an important indicator of atmospheric circulation, wind in these bands is influenced by the Hadley and Walker circulations, which are the mean meridional and longitudinal overturning circulations, respectively. In particular, the Walker circulation is affected by the El Ni\~{n}o-Southern Oscillation (ENSO) over the Pacific Ocean \citep{gastineau2009hadley}. Therefore, for $-30^{\circ}< L_m < 30^{\circ}$ we assume that (${\xi}_{L_{m}}$, ${\tau}_{L_{m}}$) are fixed and equal to the estimated value from the adjacent band fit in Figure \ref{fig:var1}, whereas we assume a constant value equal to (${\xi}$, ${\tau}$) outside this range and ($a$,$b$) for all latitudinal bands. The parameter estimates and corresponding $95\%$ confidence intervals are $\hat{a}=0.136$ $(0.132, 0.140)$, $\hat{b}=0.071$ $(0.067, 0.075)$, $\hat{\xi}=0.960$ $(0.903, 1.000)$ and $\hat{\tau}=0.628$ $(0.626, 0.630)$. The latitudinal parameters are then ${\bm \theta}_{\rm lat}=(a,b, {\xi}_{L_{m}}, {\tau}_{L_{m}})^{\top}$ for $m$ such that the latitudes are in the range of $-30^{\circ}< L_m < 30^{\circ}$. They are otherwise constant.

\subsection{Inference}\label{sec:fit}

A computational benefit of axially symmetric models on regularly spaced data is that the resulting covariance matrix is block circulant and hence block diagonal in the spectral domain \citep{davis1979circulant}. Thus, likelihood evaluation is convenient in the spectral domain, requiring matrix inversion and determinant computation of small matrices \citep{jun2008nonstationary}. In case of a nonstationary model across longitude at a given latitude, it is still possible to derive a likelihood expression whose computational efficiency is close to that of the axially symmetric case if the data are on a regular grid. 

Let $\bm{\theta}=(\bm{\theta}_{\rm time}^{\top},\bm{\theta}_{\rm lon}^{\top},\bm{\theta}_{\rm lat}^{\top})^{\top}$, where $\bm{\theta}_{\rm time}$, $\bm{\theta}_{\rm lon}$, and $\bm{\theta}_{\rm lat}$ are collections of all temporal, longitudinal, and latitudinal parameters, respectively. If the data are on a grid, \eqref{eq:likelihood1} simplifies to
\be
2l({\bm\theta};{\bf D})&=&TNM(R-1)\log(2\pi)+TNM\log(R)\nonumber\\
&&+(R-1)\sum_{m=1}^{M}\log|{\bm \Sigma}_{m}^{1}({\bm\theta}_{\rm lon})|+(R-1)\sum_{p=1}^{P}\log|{\bm \Sigma}_{p}^{2}({\bm\theta}_{\rm lat})|\label{eq:likelihood2}\\
&&+\sum_{r=1}^{R}\sum_{k=1}^{K}\sum_{p=1}^{P}\bold{v}_{p}(t_{k},r;\bm{\theta}_{\rm time},\bm{\theta}_{\rm lon})^{\top} {\bm \Sigma}_{p}^{2}({\bm\theta}_{\rm lat})^{-1} \bold{v}_{p}(t_{k},r;\bm{\theta}_{\rm time},\bm{\theta}_{\rm lon})\nonumber,
\ee
where ${\bm \Sigma}_{m}^{1}({\bm\theta}_{\rm lon})$ is the $N\times N$ coherence matrix of latitudinal band $L_{m}$, ${\bm \Sigma}_{p}^{2}({\bm\theta}_{\rm lat})$ is the $\left(M\times \lfloor N/P\rfloor\right) \times \left(M\times \lfloor N/P\rfloor\right)$ covariance matrix describing the coherence among multiple latitudinal bands, which is obtained by approximating $\bm{\varphi}_{L_{m}}$ in \eqref{phisigma_var} with $p=1,\ldots,P$ diagonal blocks, and the vector $\bold{v}_{p}(t_{k},r;\bm{\theta}_{\rm time},\bm{\theta}_{\rm lon})$ is a suitable transformation of $\bold{D}$ \citep{castruccio2014beyond}. To estimate the spatial and temporal structure of the data, we use \eqref{eq:likelihood2} throughout this study. 

As ${\bm\theta}$ is typically very high dimensional, we achieve an approximate maximum likelihood estimator by applying \eqref{eq:likelihood2} under a conditional approximations inference scheme that assumes independence across increasingly large subsets, as in \cite{castruccio2013global}.  Each approximation assumes that the parameters obtained from previous steps are fixed and known for the upcoming steps: 
\begin{itemize}
\itemsep0em
\item[] Step 1. Estimate the temporal parameters, $\bm{\theta}_{\rm time}$, by assuming that there is no cross-temporal dependence in latitude and longitude;
\item[] Step 2. Consider that $\bm{\theta}_{\rm time}$ is fixed at its estimated value and estimate $\bm{\theta}_{\rm lon}$ by assuming that the latitudinal bands are independent;
\item[] Step 3. Consider $\bm{\theta}_{\rm time}$ and $\bm{\theta}_{\rm lon}$ fixed at their estimated values and estimate $\bm{\theta}_{\rm lat}$. 
\end{itemize}
Since steps 1 and 2 assume independence across subsets, inference can be performed independently by multiple processors in a workstation or in a cluster. 

As argued by \cite{castruccio2017evolutionary}, the sequential approach with previously estimated parameters could produce an estimation bias. This is mostly apparent from step 2 to 3, where the estimated parameters for the single latitudinal band approximation may not be the optimal values for the multiple latitudinal band approximation. One solution to mitigate this issue is to refit $\bm{\theta}_{\rm lon}$ for two adjacent bands. This step requires additional computational time, 1.5 to 2 hours on a 24-cores workstation for the ALT-VAR model (parallelizing the inference for different sets of contiguous bands) but it improved model fit markedly in this study. This can be done for several adjacent bands if the computational time is acceptable, but refitting all bands with the full data set may require several weeks of computational time and very powerful computational resources.

\section{Model Comparison and Validation of Local Behavior}\label{sec:comparison}

We compare the model introduced in the previous section with previously available models, and we validate the local space-time structure against the data. 

Table~\ref{tab1} presents a comparison in terms of model selection metrics: a land/ocean evolutionary spectrum with a nonstationary latitudinal AR(1) process (LAO-AR(1)), our new evolutionary spectrum with a nonstationary latitudinal AR(1) process (ALT-AR(1)) and with a nonstationary latitudinal VAR(1) process (ALT-VAR(1)). ALT-AR(1) requires approximately 1.67 times more parameters than does LAO-AR(1), but it shows clear improvements in terms of the normalized log-likelihood, BIC and other standard model selection metrics (not shown). ALT-AR(1) allows for spatially varying coefficients across the mountain profiles and shows a noticeable improvement in model fit as the log-likelihood improves by 0.08 units per observation. The most general ALT-VAR(1) requires two additional parameters $a$ and $b$, and it achieves a further improvement in the fit. While the relative improvement between ALT-VAR(1) and ALT-AR(1) compared to the improvement between LAO-AR(1) and ALT-AR(1) is not conspicuous, the results in Table~1 are expressed in $10^8$ units and, as Figure \ref{fig:var1}(a) highlights, the improvement in absolute terms is far from being negligible: the BIC improves hundreds, or even thousands of units in some latitudes. 

\begin{table}[htbp]\centering
\caption{\label{tab1}Comparison between different models in terms of the number of parameters (excluding the temporal component), the normalized restricted log-likelihood, and BIC. The general guidelines for $\Delta$loglik$/\{NMT(R-1)\}$ are that anything above 0.1 is large and anything above 0.01 is modest but still sizable \citep{castruccio2013global}.}
\begin{tabular}{|c||c|c|c|c|}
\hline
Model  &  LAO-AR(1) &  ALT-AR(1) & ALT-VAR(1)\\
\hline
\# of parameters & $1202$ & $2006$ & $2008$\\
\hline
$\Delta$loglik$/\{NMT(R-1)\}$ & $0$ & $0.08152$ & ${\bf 0.08177}$ \\
\hline
BIC ($\times 10^8)$ &  $-1.02638$ & $-1.05015$ & ${\bf -1.05023}$\\
\hline
\end{tabular}
\end{table}

We assess the high-frequency behavior of the models by computing the contrast variances to assess the quality of the fit \citep{jun2008nonstationary}:
\be\label{delta_con}
\begin{split}
\Delta_{ew;m,n}&=&\frac{1}{KR}\sum_{k=1}^{K}\sum_{r=1}^{R}\{{H}_{r}(L_{m},\ell_{n},t_{k})-{H}_{r}(L_m,\ell_{n-1},t_{k})\}^{2},\\
\Delta_{ns;m,n}&=&\frac{1}{KR}\sum_{k=1}^{K}\sum_{r=1}^{R}\{{H}_{r}(L_{m},\ell_{n},t_{k})-{H}_{r}(L_{m-1},\ell_{n},t_{k})\}^{2},
\end{split}
\ee
where $\Delta_{ew;m,n}$ and $\Delta_{ns;m,n}$ denote the east-west and north-south contrast variances, respectively.

\begin{figure}[!ht]\centering
\includegraphics[height=1.3in,width=2.4in]{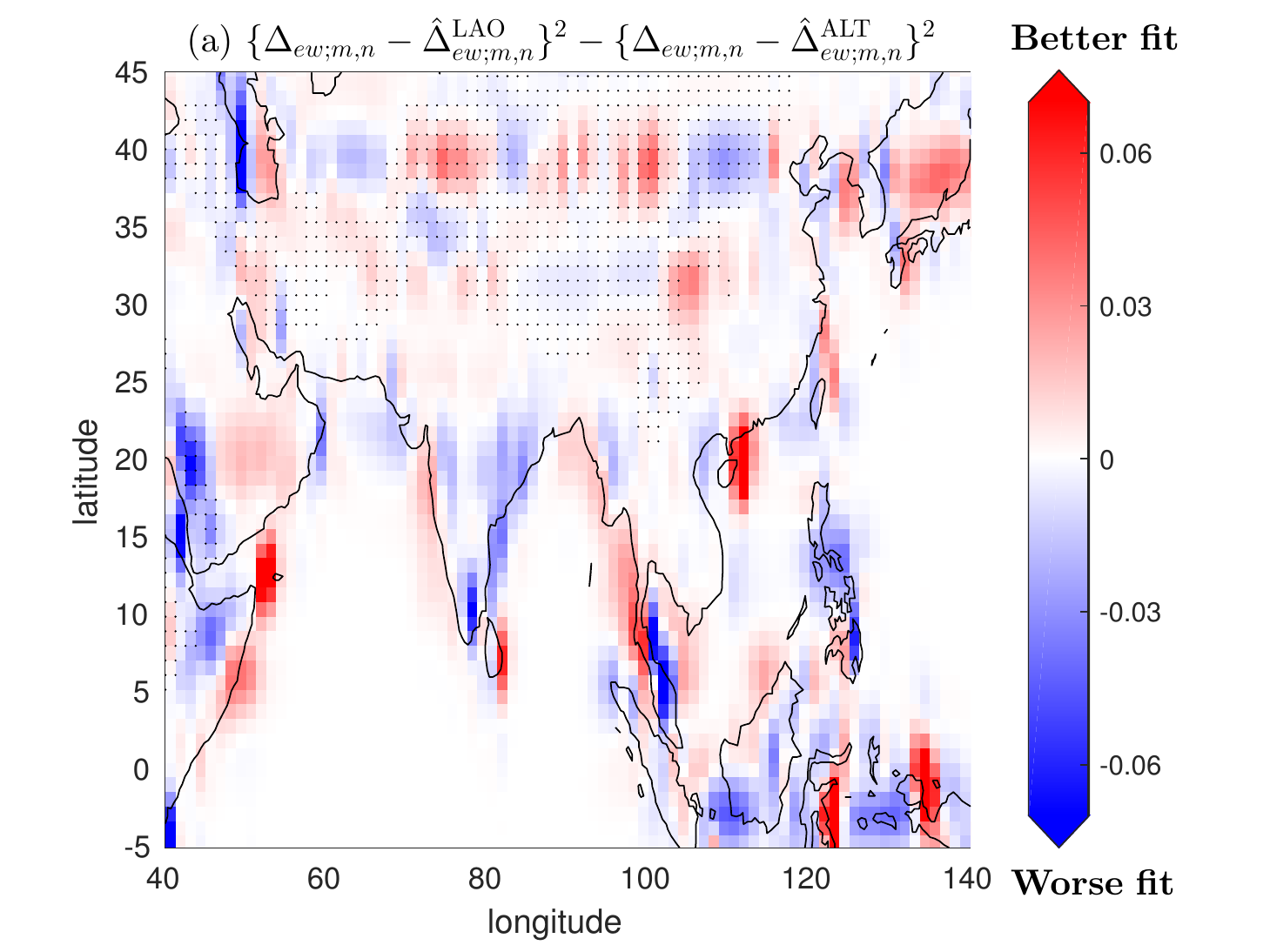}\hspace{-0.5cm}
\includegraphics[height=1.3in,width=2.4in]{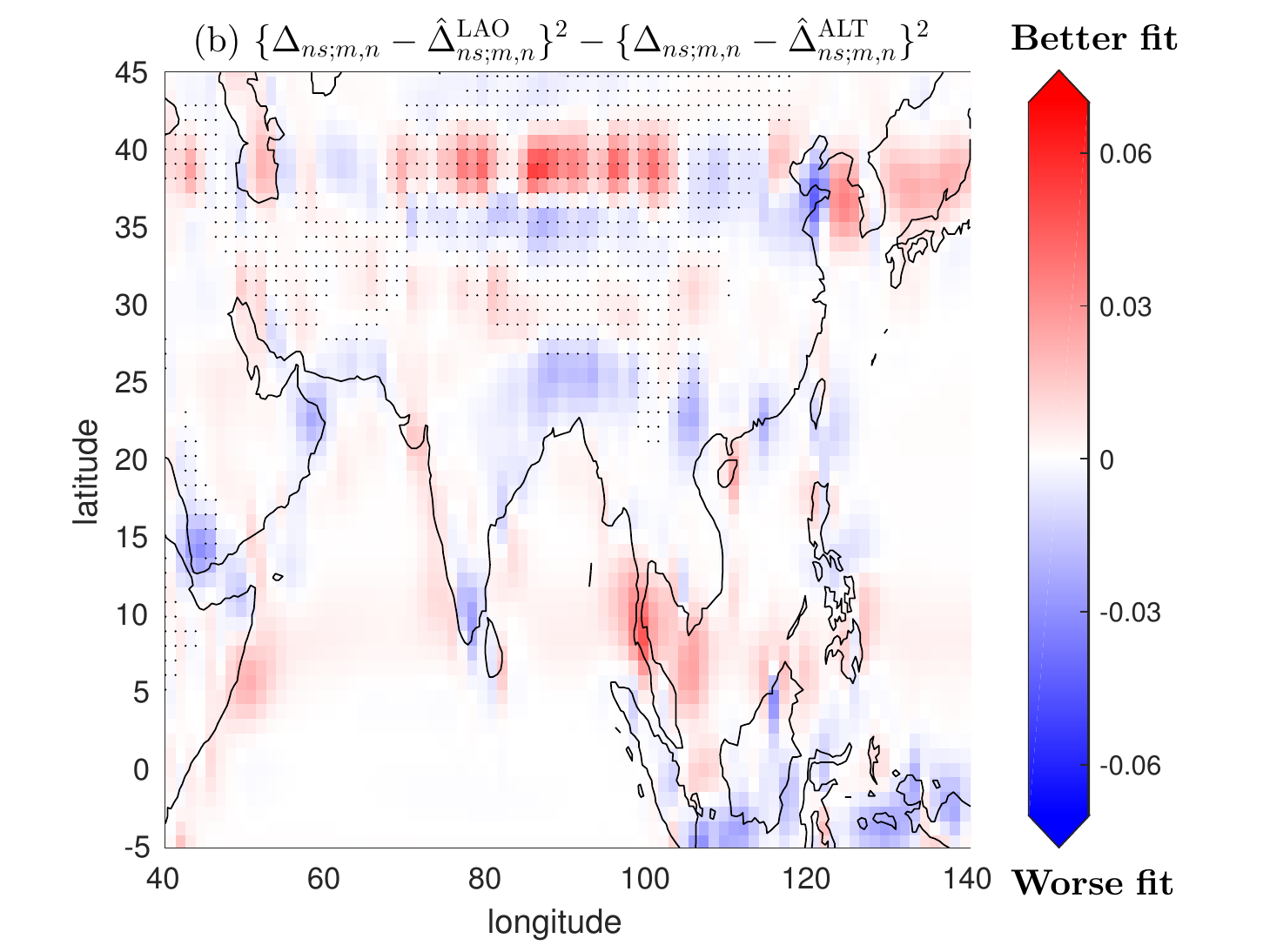}
\caption{The squared distances of the fitted contrast variances from the empirical contrast variances between two models, LAO-AR(1) and ALT-VAR(1): (a) $\{ \Delta_{ew;m,n} -\hat{\Delta}_{ew;m,n}^{\rm LAO} \}^{2} - \{ \Delta_{ew;m,n}-\hat{\Delta}_{ew;m,n}^{\rm ALT} \}^{2}$ and (b) $\{ \Delta_{ns;m,n}-\hat{\Delta}_{ns;m,n}^{\rm LAO} \}^{2} - \{ \Delta_{ns;m,n}-\hat{\Delta}_{ns;m,n}^{\rm ALT} \}^{2}$. Black dots indicate the locations where the surface altitude is larger than 1,000~m. }\label{fig:contr_diff}
\end{figure}

We compute the squared distances between the empirical and fitted variances for both LAO-AR(1) and ALT-VAR(1), and plot their differences in Figure~\ref{fig:contr_diff}. Positive and negative values represent better and worse model fit of ALT-VAR(1) compared to LAO-AR(1), respectively. The Himalayan region (from $78.75^{\circ}$E to $86.25^{\circ}$E and from $26.86^{\circ}$N to $30.63^{\circ}$N) has considerably more positive values for the north-south contrast variance case in Figure~\ref{fig:contr_diff}(b). It is also apparent how ALT-VAR(1) shows a better model fit near the Tian Shan mountain region (from $72.5^{\circ}$E to $80^{\circ}$E and from $38.16^{\circ}$N to $41^{\circ}$N) with positive values for both east-west and north-south contrast variance cases.

To quantify the improvement corresponding to these mountain ranges, we computed the aforementioned difference among these two mountain regions and compared their distributions. Table~\ref{tab:diff} represents the $25$th, $50$th, $75$th percentiles of difference near Himalayan and Tian Shan mountain regions, and we observe that overall the distributions tend to have more positive values, i.e., ALT-VAR(1) has better model fit in terms of contrast variances compared to LAO-AR(1). The table also confirms the visual inspection in Figure~\ref{fig:contr_diff}: the two metrics have larger values near Tian Shan mountain region compared to near Himalayan region.

\begin{table}[htbp]\centering
\caption{\label{tab:diff}$25$th, $50$th and $75$th percentiles of two difference metrics near Himalayan region (H) and Tian Shan mountain region (T).}
\setlength\tabcolsep{4.5pt}
\begin{tabular}{|c||c||c|c|c|}
\hline
Metric  & Region & $25$th & $50$th & $75$th \\
\hline
\multirow{2}{*}{$[ \{ \Delta_{ew;m,n}-\hat{\Delta}_{ew;m,n}^{\rm LAO} \}^{2} - \{ \Delta_{ew;m,n}-\hat{\Delta}_{ew;m,n}^{\rm ALT} \}^{2} ]\times 10^3$ }  & H & $-1$ & $1$ & $2$ \\
& T & $-9$ & $20$ & $57$ \\
\hline
\multirow{2}{*}{$[ \{ \Delta_{ns;m,n}-\hat{\Delta}_{ ns;m,n}^{\rm LAO} \}^{2} - \{ \Delta_{ ns;m,n}-\hat{\Delta}_{ ns;m,n}^{\rm ALT} \}^{2} ]\times 10^3$} & H & $0$ & $6$ & $10$ \\
& T & $-3$ & $8$ & $52$ \\
\hline
\end{tabular}
\end{table}

\section{Generation of Stochastic Surrogates and Validation of Large-Scale Behavior}\label{sec:simul}

In this section, we explain how to generate the stochastic surrogates from the SG. Besides their interest for wind energy assessment, such surrogate runs can then be compared with the original LENS runs to validate the large-scale behavior of the statistical model.

In the previous sections $\bm{\theta}=(\bm{\theta}_{\rm time}^{\top},\bm{\theta}_{\rm lon}^{\top},\bm{\theta}_{\rm lat}^{\top})^{\top}$ in \eqref{mod_ensemble} have been defined and estimated from the training set. The mean climate ${\bm \mu}$ can be obtained as a smoothed version of the ensemble mean $\overline{\bf W}$. Similarly to \cite{castruccio2016compressing} and \cite{castruccio2017evolutionary}, for each location $(L_m,\ell_n)$ we fit a smoothing spline $\widetilde{W}(L_{m},\ell_{n},t_{k})$ for $k=1,\ldots, K$, which minimizes 
\be
\lambda \sum_{k=1}^{K}\Big\{\overline{W}(L_{m},\ell_{n},t_{k})-{\widetilde{W}}(L_{m},\ell_{n},t_{k})\Big\}^{2}+(1-\lambda)\sum_{k=1}^{K} \Big\{\nabla_2 \widetilde{W}(L_{m},\ell_{n},t_{k}) \Big\}^{2}\nonumber,
\ee
where $\nabla_2$ is the second-order finite difference operator. We impose a penalty term, $\lambda=0.01$, to reflect the slowly varying climate of annual wind fields over the next century \citep{vaughan2013remote}. 

Once ${\bm \mu}$ and $\bm{\theta}$ are estimated, surrogate runs can be almost instantaneously generated on a modest laptop by performing the following steps:
\begin{itemize}
\itemsep0em
\item[] Step 1. Generate ${\bold{e}}_{L_{m}}\stackrel{\rm iid}{\sim}\mathcal{N}(\bold{0},{\bold{\Sigma}}_{L_{m}})$ as in \eqref{eq:var1};
\item[] Step 2. Compute $\widetilde{\bold{H}}_{L_{m}}$ with expressions \eqref{eq:var1};
\item[] Step 3. Compute ${H}_{r}(L_{m},\ell_{n},t_{k})$ with expression \eqref{eq:gft};
\item[] Step 4. Compute ${\bm \epsilon}_{r}$ with equation \eqref{eq:ar2};
\item[] Step 5. Obtain the reproduced run as $\widetilde{\bold W}+{\bm \epsilon}_{r}$, where 
\end{itemize} 
\[
\widetilde{\bf W}=\{\widetilde{W}(L_{1},\ell_{1},t_{1}),\dots,\widetilde{W}(L_{M},\ell_{1},t_{1}),\widetilde{W}(L_{1},\ell_{2},t_{1}),\dots,\widetilde{W}(L_{M},\ell_{N},t_{K})  \}^{\top}.
\]

\begin{figure}[htb]\centering
\includegraphics[height=1.3in,width=2.4in]{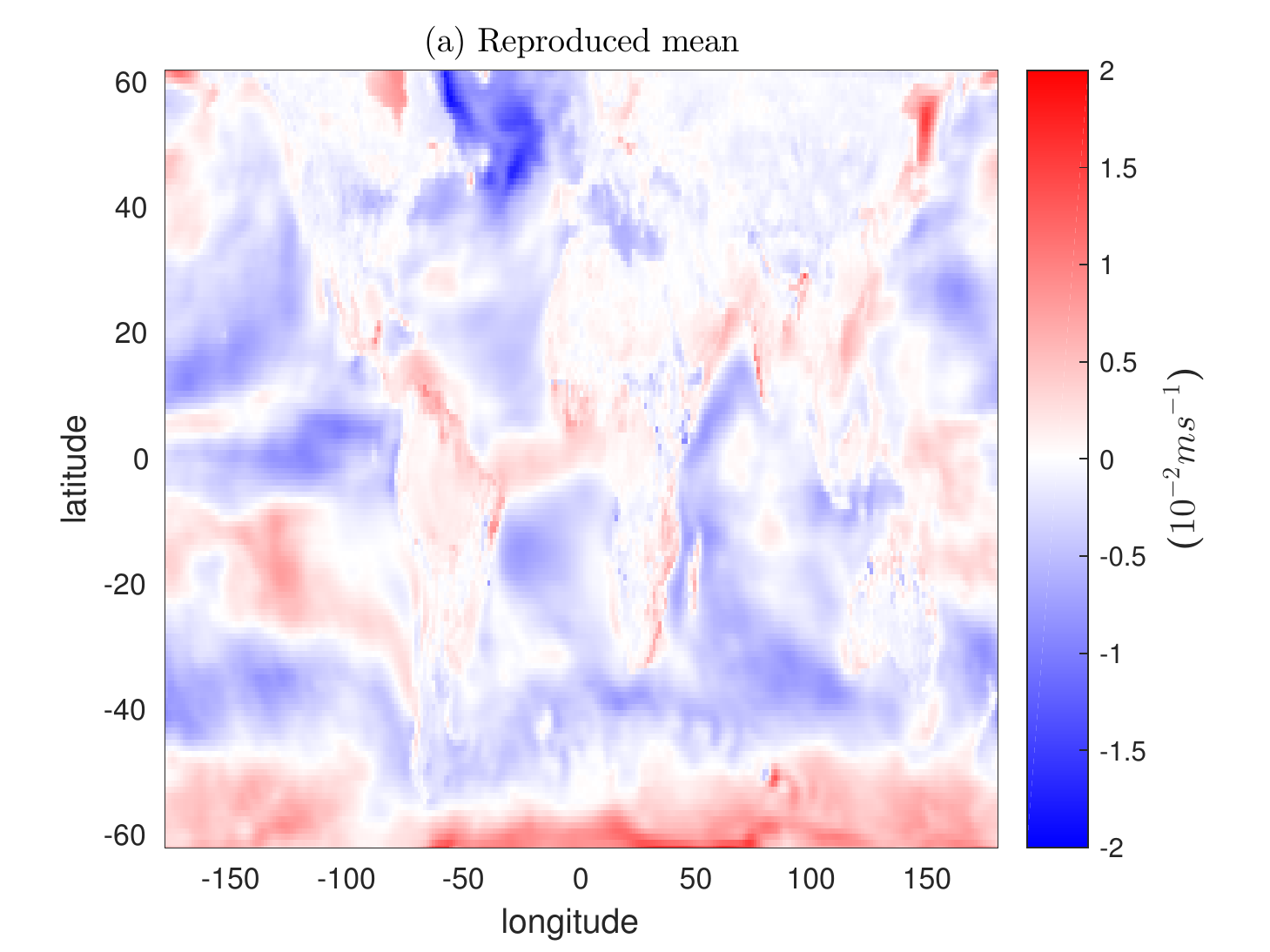}
\includegraphics[height=1.3in,width=2.4in]{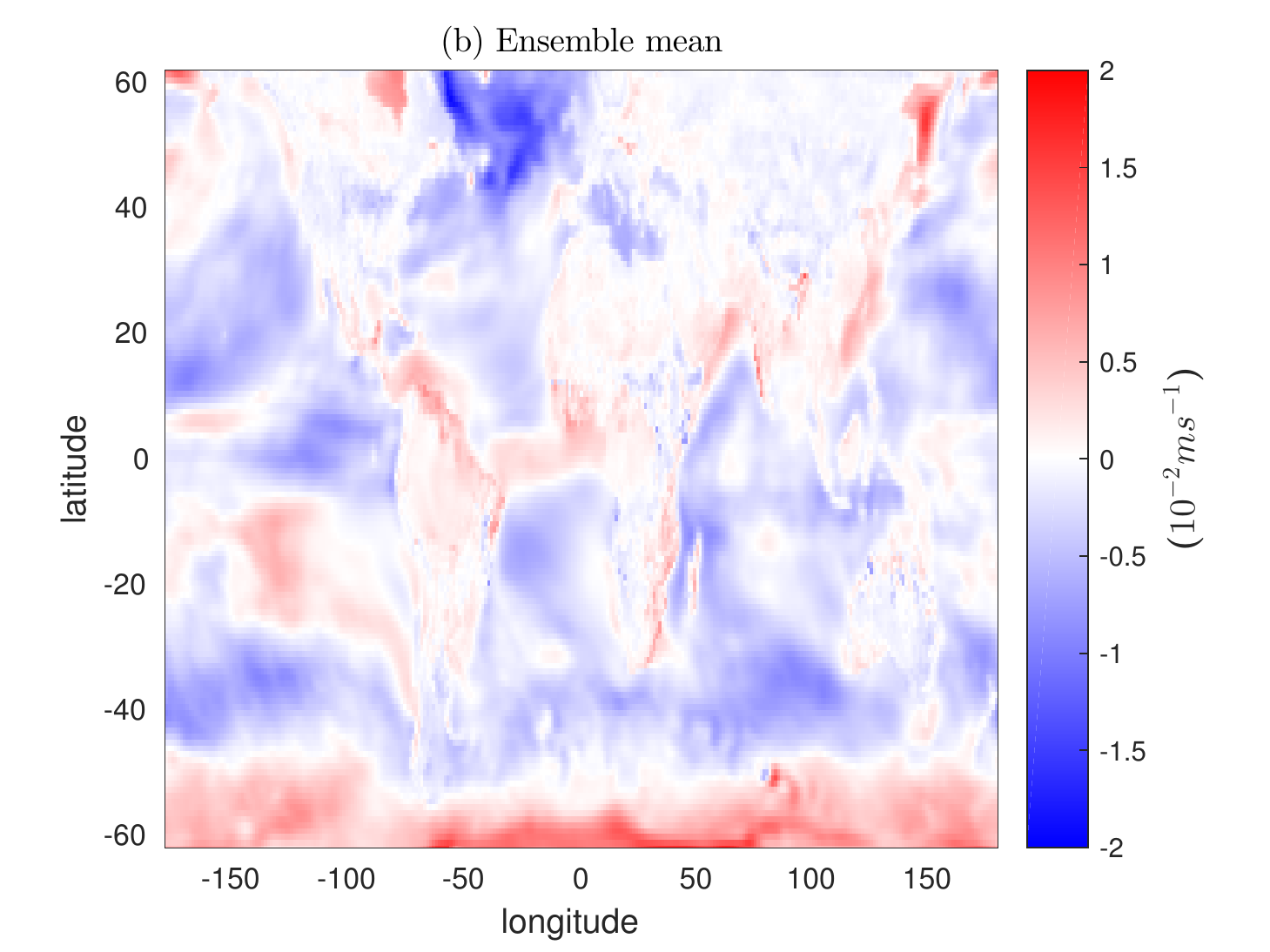}
\includegraphics[height=1.3in,width=2.4in]{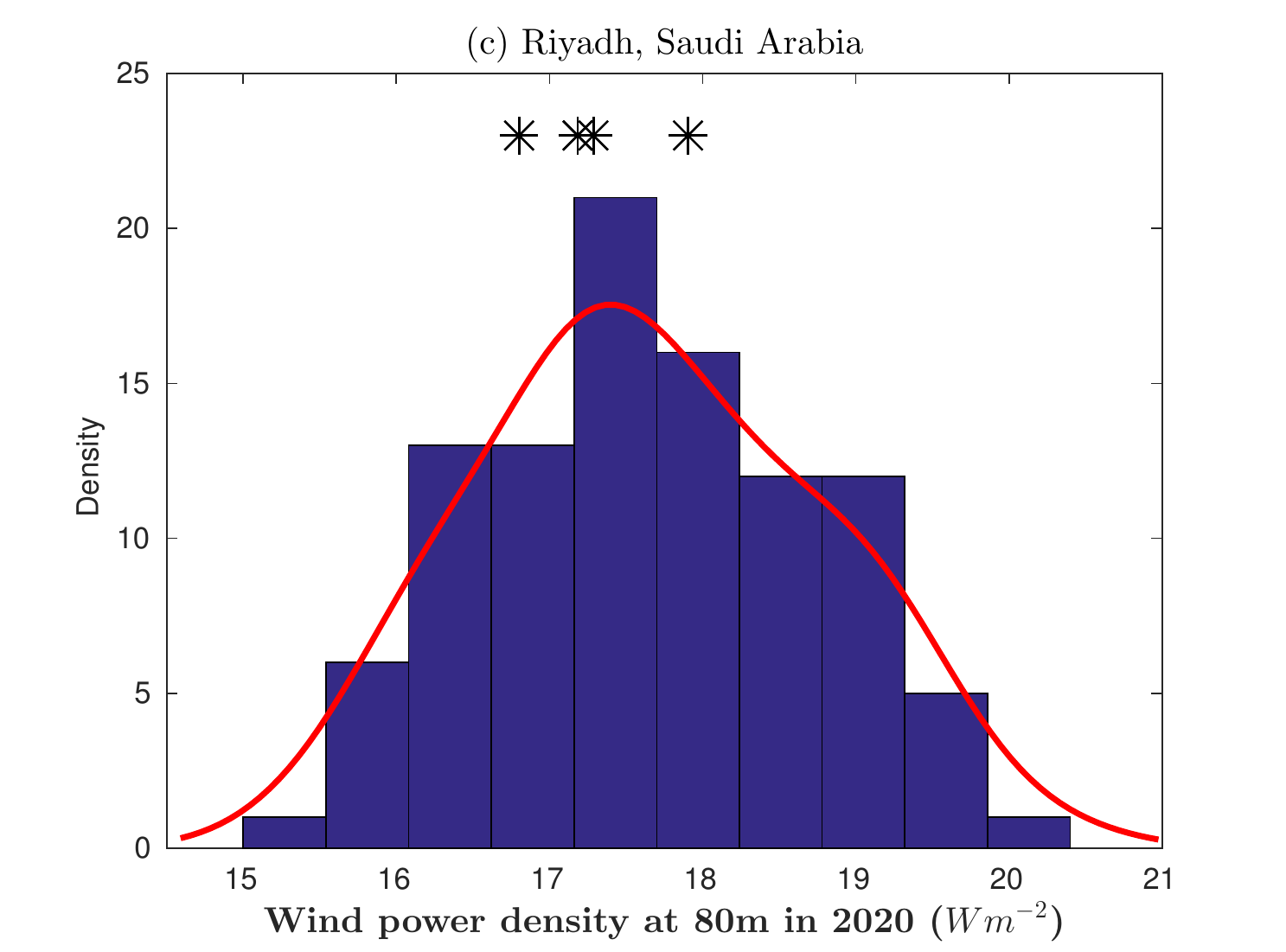}
\includegraphics[height=1.3in,width=2.4in]{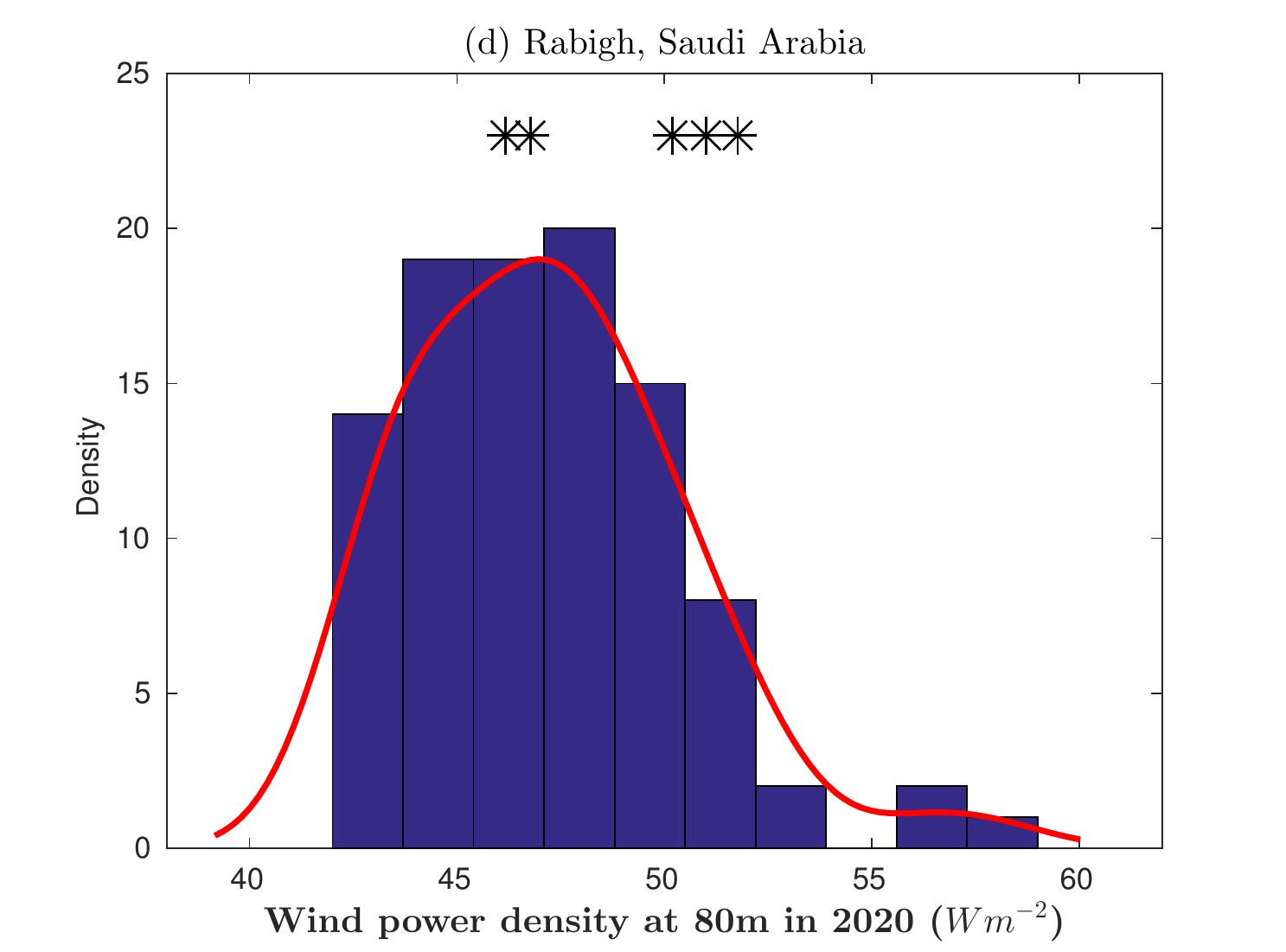}
\caption{Top: Global maps of (a) the mean from reproduced runs and (b) the ensemble mean of the near-future ($2013-2046$) annual near-surface wind speed trends. Bottom: Histogram of the distribution of the wind power density at $80$~m in $2020$ with nonparametric density in red for the one-hundred reproduced runs near (c) Riyadh and (d) Rabigh, Saudi Arabia ($\ast$ represents the original climate model runs).}\label{fig:34years_sts}
\end{figure}

We generated one hundred runs and compared them with the climate model runs, see Figure S8 for a comparison in 2050 of five runs with other five LENS runs not in the training set and a movie of a surrogate run (Movie S1). We computed near-future (2013-2046) annual wind speed trends (a reference metric in the reference LENS publication \citep{kay2015community}) for each of the surrogate and LENS runs and then plotted the corresponding means in Figures~\ref{fig:34years_sts}(a) and \ref{fig:34years_sts}(b) (see Figure S7 for a comparison of the individual runs), and the $2.5$th, $50$th, and $97.5$th percentiles in 2050 in Figure S9. From these figures, it is apparent how the SG and LENS distributions are visually indistinguishable, with a stronger trend over ocean and coastline than over land. 

Figure~\ref{fig:34years_sts}(c-d) shows a comparison between reproduced and climate model runs in terms of their distribution of wind power density at $80$~m in $2020$ (details on how to derive this variable from wind speed are provided in the supplementary material \citep{jeong2017reducingsupplement}) for locations near Riyadh ($24.97^{\circ}$N and $46.25^{\circ}$E) and Rabigh, Saudi Arabia ($23.01^{\circ}$N and $38.75^{\circ}$E). Both locations are in the Arabian peninsula and exhibit significant non-decreasing trends. So, an assessment of the internal variability is crucial to determining the robustness of the point estimates and could inform policy makers on the uncertainty and associated risks in building wind turbines in these areas where no regional studies and very limited ground-based observations are available. Here, we observe that Rabigh on the coastline has considerably more potential to generate wind power than Riyadh in the central inland of Saudi Arabia. A more accurate assessment of wind resources could be achieved by using wind speed data at a  higher spatio-temporal resolution than the one used in this study (i.e., annual mean wind speed at horizontal resolution of approximately $1^{\circ}$), but such an assessment is currently unfeasible given the absence of ESM simulations at fine spatio-temporal resolutions for multiple decades. The five climate model runs are poorly informative for internal variability, but the distribution generated from many reproduced runs allows for a more accurate assessment. Both locations exhibit a considerable variability in wind power density (2.5 and 97.5 percentiles), with (15.7, 19.7)$Wm^{-2}$ for Riyadh and (42.3, 55.9)$Wm^{-2}$ for Rabigh.

Figure \ref{fig:34years_sts}(c-d) depends only on the marginal wind at two given locations, so it could be obtained with simpler pointwise approaches without assuming spatial dependence. The SG, however, allows to generate spatially resolved fields, which are indistinguishable from the original LENS runs (see Figures S7 and S8). To visualize this interactively, a dynamic Graphical User Interface (GUI) application in Matlab is provided in the supplementary material \citep{jeong2017reducingsupplement}. The GUI requires to download $\hat{\bm \mu}$ and $\hat{\bm \theta}$ in \eqref{mod_ensemble}, for a total of 30 megabytes, instead of downloading the entire climate model ensemble (40 members), which is 1.1 gigabytes. A user can then use the stored coefficients and generate many runs to achieve a considerably more detailed assessment of wind uncertainty under different initial conditions.

\section{Discussion and Conclusion}\label{sec:concl}
 
Understanding the spatio-temporal variability of wind resources is essential to sustain the increasing energy demand, but traditional ESM ensemble-based approaches for assessment in developing countries are increasingly computationally, time and memory consuming. SGs provide a simple and computationally convenient tool for generating surrogate runs under different initial conditions and assessing the uncertainty from internal variability without storing a prohibitive amount of information. Once inference is performed and the parameters have been estimated from a small number of LENS members, an end user can download a small software package and use it to almost instantaneously generate many reproduced runs whose large-scale features are almost identical to the original runs (see Figures~\ref{fig:34years_sts}(a) and (b)) and assess the uncertainty in future wind power density due to internal variability (see Figure~\ref{fig:34years_sts}(c) and (d)). 

We introduced a spectral model for gridded data which allows for an improved fit of global wind data. Our proposed model presents two elements of novelty from the current literature:

\begin{enumerate} 

\item It incorporates more large-scale geographical information to explain the nonstationary behavior of wind across longitude. In particular, the model incorporates orography, which is shown to affect the spatial smoothness of wind fields. The proposed model allows for spatially varying parameters depending on the surface altitude over land and high mountains, contains the axially symmetric and the land/ocean evolutionary spectrum as special cases, and shows improved performance in terms of the log-likelihood, BIC and other standard model selection metrics. 

\item It introduces a nonstationary VAR(1) model for the latitudinal coherence for multiple wavenumbers. By assuming independent partitions of the correlated innovations for neighboring wavenumbers, the proposed model still holds a convenient formulation of the log-likelihood function in~\eqref{eq:likelihood2} and further improves the model fit. 

\end{enumerate}

Inference is performed via a multi-step conditional likelihood approach, which leverages on parallel computation and achieves a fit on a data set of more than 18 million data points.

For policy making purposes, a clear limitation of our approach is the coarse time scale at which wind power density is assessed. Finer time scales require considerable modeling and face computational challenges. On the modeling side, the Gaussianity assumption has to be relaxed at higher temporal resolution and requires alternative trans-Gaussian processes, such as Tukey $g$-and-$h$ random fields \citep{xu2017tukey}. On the computational side, the already considerable data size of this application (more than 18 million data points) will be increased by more than two orders of magnitude. While clearly adding a layer of complexity to inference, the same key ingredients, namely leveraging on regular geometries, parallel computing and spectral methods have already shown to achieve inference from data sets larger than one billion data points \citep{castruccio2016compressing}, so a global inference of daily wind power density for the entire ensemble is likely achievable with current computational architectures. If a smaller region such as Saudi Arabia is chosen, then the decrease in the number of spatial locations alleviates the computational burden to some extent, and would allow to model non-Gaussian processes at finer scale, see \cite{tagle2017}.

\begin{supplement}[id=suppA]
  \stitle{Supplement to ``Reducing Storage of Global Wind Ensembles with Stochastic Generators"} 
  \slink[doi]{COMPLETED BY THE TYPESETTER}
  \sdatatype{.pdf}
  \sdescription{Further technical details and a Graphical User Interface application in Matlab can be found in the online supplementary material.}
\end{supplement}

\bibliography{papers1}
\bibliographystyle{Chicago}
  
\end{document}